\documentclass[12pt]{article}
\pdfoutput=1
\usepackage{amsmath}
\usepackage{epsf}
\usepackage{cite}
\usepackage{color}
\usepackage{graphicx}

\newcommand{\beq}{\begin{equation}}
\newcommand{\eeq}{\end{equation}}
\newcommand{\be}{\begin{equation}}
\newcommand{\ee}{\end{equation}}
\newcommand{\ba}{\begin{array}}
\newcommand{\ea}{\end{array}}
\newcommand{\beqa}{\begin{eqnarray}}
\newcommand{\eeqa}{\end{eqnarray}}
\newcommand{\bea}{\begin{eqnarray}}
\newcommand{\eea}{\end{eqnarray}}
\newcommand{\beqn}{\begin{eqnarray}}
\newcommand{\eeqn}{\end{eqnarray}}

\newcommand{\D}{\Delta}

\newcommand{\nn}{\nonumber}

\newcommand{\gev}{{\rm GeV}}

\newcommand{\cO}{{\cal O}}

\newcommand{\cL}{{\cal L}}

\newcommand{\cH}{{\cal H}}

\newcommand{\lsim}{\stackrel{<}{_\sim}}

\newcommand{\epsK}{\varepsilon_K}

\definecolor{red}{cmyk}{0,1,1,0.4}
\newcommand{\hatV}{{\hat V}}
\newcommand{\hatFC}{{\hat \lambda_{\rm FC}}}

\begin{document}

\begin{flushright}
  TUM-HEP-761/10 \\
  MPP-2010-57
\end{flushright}

\medskip

\begin{center}
{\Large \bf
Higgs-mediated FCNCs: Natural Flavour \\ [0.3 cm]
 Conservation vs.~Minimal Flavour Violation
}\\[0.8 cm]
{\large Andrzej~J.~Buras$^{a,b}$, Maria Valentina Carlucci$^a$,\\
Stefania Gori$^{a,c}$, Gino Isidori$^{b,d}$} \\[0.5 cm]
\small
$^a${\em Physik-Department, Technische Universit\"at M\"unchen, James-Franck-Stra{\ss}e,
\\D-85748 Garching, Germany} \\[0.1cm]
$^b${\em TUM Institute for Advanced Study, Technische~Universit\"at~M\"unchen, Arcisstra{\ss}e 21,
\\D-80333 M\"unchen, Germany} \\[0.1cm]
$^c${\sl Max-Planck-Institut f{\"u}r Physik (Werner-Heisenberg-Institut), \\
D-80805 M{\"u}nchen, Germany}\\[0.1cm]
$^d${\em INFN, Laboratori Nazionali di Frascati, Via E.~Fermi 40, I-00044 Frascati, Italy} \\[0.8 cm]
\end{center}

\abstract{%
\noindent 
We compare the effectiveness of two hypotheses,  
Natural Flavour Conservation (NFC) and Minimal Flavour Violation (MFV), 
in suppressing the strength of  flavour-changing neutral-currents (FCNCs)
in models with more than one-Higgs doublet. We show that the MFV 
hypothesis, in its general formulation, is more stable 
in suppressing FCNCs than the hypothesis of NFC alone
when quantum corrections are taken into account.
The phenomenological implications of the two scenarios are discussed 
analysing meson-antimeson mixing observables and the rare decays $B_{s,d}\to \mu^+\mu^-$. 
We demonstrate that, introducing {\it flavour-blind} CP phases,
two-Higgs doublet models respecting the MFV hypothesis can accommodate a  
large CP-violating phase in $B_s$ mixing, as hinted by CDF and D0 data and, without extra 
free parameters, soften significantly in a correlated manner
 the observed anomaly
in the relation between $\epsK$ and $S_{\psi K_S}$.}

\section{Introduction}

The standard assignment of the $SU(2)_L\times U(1)_Y$ quark charges, 
identified long ago by Glashow, Iliopoulos, and Maiani (GIM)~\cite{Glashow:1970gm},
forbids tree-level flavour-changing couplings of the quarks to the 
Standard Model (SM) neutral gauge bosons. 
In the case of only one-Higgs doublet, 
namely within the SM, this structure is effective also 
in eliminating a possible dimension-four flavour-changing neutral-current 
(FCNC) coupling of the quarks to the Higgs field. 
While the $SU(2)_L\times U(1)_Y$ assignment of quarks and leptons 
can be considered as being well established, much less is known 
about the Higgs sector of the theory. In the presence 
of more than one-Higgs field the appearance of tree-level FCNC 
is not automatically forbidden by the standard assignment 
of the $SU(2)_L\times U(1)_Y$ fermion charges: additional conditions have to 
be imposed on the model in order to guarantee a sufficient suppression of 
FCNC processes~\cite{Glashow:1976nt,Paschos:1976ay}. 
The absence of renormalizable couplings contributing 
at the tree-level to FCNC processes, in multi-Higgs 
models, goes under the name of Natural Flavour Conservation (NFC)
hypothesis.

The idea of NFC has been with us for more than 30 years. 
During the last decade another concept for the suppression 
of FCNC processes has become very popular: the hypothesis of
Minimal Flavour Violation (MFV)~\cite{Buras:2000dm,D'Ambrosio:2002ex},
whose origin, in specific new-physics (NP) models, can be traced back 
to~\cite{Chivukula:1987py,Hall:1990ac}. The question then arises 
how NFC (and GIM) are related to MFV, and vice versa. 
Motivated by a series of recent studies about the strengths of 
FCNCs in multi-Higgs doublet 
models~\cite{Joshipura:2007cs,Pich:2009sp,Botella:2009pq,Gupta:2009wn,Ferreira:2010xe}, 
in this paper we present 
a detailed analysis of the relation between the NFC and MFV  
hypotheses.
As we will show, while the two hypotheses 
are somehow equivalent at the tree-level, important differences 
arise when quantum corrections are included. Beyond the tree
level, or beyond the implementation of these two hypotheses in
their simplest version, some FCNCs are naturally generated
in both cases. In this more general framework, the MFV hypothesis
in its general formulation~\cite{D'Ambrosio:2002ex}
turns out to be more stable in suppressing FCNCs than 
the hypothesis of NFC alone. 

This analysis will also give us the opportunity to compare 
the various formulations of  MFV present in the literature and 
to clarify which of the multi-Higgs models proposed 
in~\cite{Joshipura:2007cs,Pich:2009sp,Botella:2009pq}
are consistent with the MFV principle, and thus are naturally 
protected against too large FCNCs. 
 
The phenomenological tests of these different concepts which 
can be obtained on the basis of meson-antimeson mixing observables,
such as the CP-violating (CPV) observable $\epsK$, the mass differences $\Delta M_{d,s}$, 
and the CP asymmetries $S_{\psi K_S}$ and $S_{\psi\phi}$ are also analysed.
Beside being perfectly consistent with present data even for light  
Higgs boson masses, two-Higgs doublet models respecting 
the MFV hypothesis could even accommodate a large CP-violating phase 
in $B_s$ mixing, as hinted by CDF~\cite{Aaltonen:2007he} 
and D0~\cite{Abazov:2008fj,Abazov:2010hv}.
However, as pointed out first in~\cite{Kagan:2009bn},
this can happen only introducing flavour-blind phases, i.e.~decoupling 
the breaking of the flavour group from the breaking of the 
CP symmetry~\cite{Kagan:2009bn,Mercolli:2009ns,Paradisi:2009ey}.
We demonstrate that, introducing flavour-blind CPV phases, 
such models\footnote{The concrete two-Higgs doublet model belonging to 
this class will be called $\text{2HDM}_{\overline{\text{MFV}}}$ with the ''bar''
signalling the presence of flavour-blind CPV phases.} are not only capable of accommodating 
a large CPV phase in $B_s$ mixing: also the observed anomaly 
in the relation between $\epsK$ and $S_{\psi K_S}$~\cite{Lunghi:2008aa,Buras:2008nn}
is substantially softened
in a strictly correlated manner. 
We finally stress the key role of $B_{s,d}\to \ell^+\ell^-$ decays 
in providing a future clean experimental tests of the MFV  
hypothesis in the Higgs sector, independently of possible 
 flavour-blind phases.

The paper is organized as follows. In Section~\ref{sec:Yukawa} 
we define the two hypotheses of NFC and MFV starting from the 
general quark Yukawa coupling with two-Higgs doublets.
In Section~\ref{sec:NFC} we analyse the problems of implementing 
NFC beyond the tree-level. The stability of  MFV beyond the 
lowest order, and the comparison with the previous literature, 
is presented in Section~\ref{sec:MFV} and~\ref{sec:comparison}, respectively.
The phenomenological tests of $\text{2HDM}_{\overline{\text{MFV}}}$ by means of $\epsK$, 
$\Delta M_{s,d}$,  $S_{\psi K_S}$, $S_{\psi\phi}$, and  $B_{s,d}\to \ell^+\ell^-$decays
are discussed in Section~\ref{sec:pheno}.
We close our paper with a list of main lessons obtained through
our analysis. Some technical details on the Higgs potential and our notations can be 
 found in an Appendix.

\section{NFC and MFV hypotheses: definition and implementation
to lowest order}
\label{sec:Yukawa} 

Let's consider a model with two-Higgs doublets, $H_1$ and 
$H_2$, with hypercharges $Y=1/2$ and $Y=-1/2$, respectively. 
The most general renormalizable and
gauge-invariant interaction of these fields with the SM quarks is 
\begin{eqnarray}
- \cL_Y^{\rm gen} = \bar Q_L X_{d1} D_R H_1 + \bar Q_L X_{u1} U_R H_1^c 
+ \bar Q_L X_{d2} D_R H_2^c + \bar Q_L X_{u2} U_R H_2 +{\rm h.c.}~,
\label{eq:generalcouplings}
\end{eqnarray}
where $H_{1(2)}^c = -i\tau_2 H_{1(2)}^*$  and the $X_i$ 
are $3\times 3$ matrices with a generic flavour structure.
The quark mass matrices are linear combinations of the  matrices
$X_i$, weighted by the corresponding 
Higgs vacuum expectation values (vevs):
\be
\label{eq:MdMu}
M_d = \frac{1}{\sqrt{2}} \left( v_1 X_{d1} + v_2 X_{d2} \right)~, \qquad 
M_u = \frac{1}{\sqrt{2}} \left( v_1 X_{u1} + v_2 X_{u2} \right)~.
\ee 
Here $\langle H^\dagger_{1(2)} H_{1(2)} \rangle=v^2_{1(2)}/2$, 
with $v^2 =v_1^2+v_2^2 \approx (246~{\rm GeV})^2$ and, 
by means of global phase transformations of $H_{1,2}$, 
we have eliminated possible CPV phases in the Higgs vevs
(i.e.~we have shifted CPV phases in the Higgs interaction terms).
For generic $X_i$ we cannot diagonalize simultaneously 
these two mass matrices and the couplings to the three 
physical neutral Higgs fields. Consequently we are left with 
dangerous FCNC couplings to some of them.

The $X_i$ break in different ways the large quark-flavour 
symmetry of the gauge sector of the SM. They also break 
possible continuous or discrete symmetries associated to 
the Higgs sector. A convenient classification of various 
two-Higgs doublet models, and of the possible protection of
FCNCs is obtained by identifying how these symmetries are broken. 
For simplicity, we focus the following discussion only on 
the quark sector of a two-Higgs doublet model (2HDM), 
but the analysis can easily be generalized to include 
the lepton sector and more than two-Higgs doublets. 

The largest group of unitary quark field transformations that 
commutes with the SM gauge Lagrangian can be decomposed as~\cite{Chivukula:1987py,D'Ambrosio:2002ex},
\be
{\mathcal G}_q =  {\rm SU}(3)^3_q \otimes {\rm U}(1)_B \otimes {\rm U}(1)_Y \otimes {\rm U}(1)_{\rm PQ}~,
\ee
where 
\be
{\rm SU}(3)^3_q = {\rm SU}(3)_{Q_L}\otimes {\rm SU}(3)_{U_R} \otimes {\rm SU}(3)_{D_R}
\ee
and the three $U(1)$ symmetries are the baryon number, the hypercharge, and 
the Peccei-Quinn symmetry~\cite{Peccei:1977ur}, respectively.
As far as ${\rm U}(1)_{\rm PQ}$ is concerned, we define it as the symmetry 
under which $D_R$ and $H_1$ have opposite charge, while all the other 
fields are neutral. Since we assume that hypercharge is not explicitly 
broken, and that baryon number is conserved, the two ingredients
in the classification of the structure of the Yukawa interaction are:
\begin{itemize}
\item the breaking of the flavour-blind $U(1)_{\rm PQ}$ symmetry 
and of other discrete flavour-blind symmetries involving both right-handed 
quarks and Higgs fields;
 \item the breaking of the ${\rm SU}(3)^3_q$
flavour symmetry.
\end{itemize}
According to which of these two breaking mechanism is {\em protected}, 
we can identify the two frameworks we are interested in:
\begin{itemize}            
\item The {\it{Natural Flavour Conservation}} hypothesis, formulated in~\cite{Glashow:1976nt}, 
is the assumption that only one-Higgs field can couple to a given quark species. 
This structure can be implemented by appropriate {\em flavour-blind} symmetries. 
In particular, the so-called type-II model, namely the condition
\be
X_{u1}= X_{d2} =0~\qquad{\rm [NFC, Type-II]}~,
\label{eq:NFCII}
\ee
is obtained requiring the invariance of $\cL_Y^{\rm gen}$ under ${\rm U}(1)_{\rm PQ}$. 
The same result can be obtained using a discrete subgroup of ${\rm U}(1)_{\rm PQ}$:
the $Z_2$ symmetry under which $H_1 \to - H_1$, $D_R \to - D_R$ and all other fields are unchanged. 
Another realization of the NFC hypothesis is the so-called type-I model,
namely the condition
\be
X_{u2}= X_{d2} =0~\qquad{\rm [NFC, Type-I]}~,
\label{eq:NFCI}
\ee
that can be obtained imposing the 
$Z_2$ symmetry under which only $H_2 \to - H_2$ and all other fields are unchanged.
\item The  {\it{Minimal Flavour Violation}} hypothesis, as 
formulated in~\cite{D'Ambrosio:2002ex}, is the assumption that 
the $SU(3)_q^3$ flavour symmetry is broken only by two independent terms,
$Y_d$ and $Y_u$, transforming as
\be
Y_u \sim (3, \bar 3,1)_{{\rm SU}(3)^3_q}~,\qquad
Y_d \sim (3, 1, \bar 3)_{{\rm SU}(3)^3_q}~.
\label{eq:MFVspur}
\ee
Expanding to the lowest non-trivial order in these breaking terms 
leads to the following structure for the $X_i$ couplings: 
\be
\ba{ll}
X_{d1}= c_{d1} Y_d \qquad  & X_{d2}= c_{d2} Y_d \\ 
X_{u1}= c_{u1} Y_u \qquad  & X_{u2}= c_{u2} Y_u 
\ea
\qquad{[{\rm MFV},~\cO(Y^1)]}~,
\label{eq:MFV0}
\ee
where the $Y_{u,d}$ are $3\times 3$ matrices and 
the $c_i$ are arbitrary (flavour-blind) coefficients.
If the breaking of the  $SU(3)_q^3$ flavour group and the 
breaking of CP are decoupled, i.e.~if we allow the introduction 
of flavour-blind phases in the MFV framework~\cite{Kagan:2009bn,Mercolli:2009ns,Paradisi:2009ey},
the $c_i$ coefficients in Eq.~(\ref{eq:MFV0}) can be complex. 
\end{itemize}

The structure in Eq.~(\ref{eq:MFV0}), with complex $c_i$, has recently been 
postulated by Pich and Tuzon in Ref.~\cite{Pich:2009sp}.  
These authors introduced this structure as an alternative to discrete symmetries
in the Higgs sector to avoid FCNCs in a general two-Higgs doublet 
model. Here we have shown that this ansatz 
can be straightforwardly derived from the MFV hypothesis 
about the breaking of the $SU(3)_q^3$ flavour group~\cite{D'Ambrosio:2002ex}, 
generalized to include flavour-blind phases~\cite{Kagan:2009bn,Mercolli:2009ns,Paradisi:2009ey},
in the limit were the expansion in the  $SU(3)_q^3$ breaking terms is
truncated to the first  order. As we will discuss in the following, 
the MFV hypothesis is a key ingredient to make this ansatz sufficiently 
stable beyond the tree-level when 
two-Higgs doublet model is considered to be only a low-energy effective theory
(as expected by naturalness arguments). 

To explicitly check that the conditions in Eq.~(\ref{eq:MFV0}) lead to the absence of FCNCs at tree-level, 
it is convenient to change  the basis for the Higgs fields, moving to 
the basis where only one-Higgs doublet has a non-vanishing vev (see Appendix).
This is achieved by the rotation
\be
\left( \ba{c} \Phi_v \\ \Phi_H \ea \right) = \left( \ba{cc} c_\beta & s_\beta \\
-s_\beta & c_\beta \ea \right)  \left( \ba{c} H_1 \\ H_2^{c} \ea \right)~,
\qquad c_\beta =\frac{v_1}{v}~, \qquad s_\beta =\frac{v_2}{v}~,
\qquad t_\beta = \frac{s_\beta}{c_\beta}~,
\label{eq:tbetadef}
\ee
such that  
$\langle \Phi_v^\dagger \Phi_v \rangle=v^2/2$ and 
$\langle \Phi_H^\dagger \Phi_H \rangle=0$. In this basis 
the Yukawa Lagrangian assumes the form 
\begin{eqnarray}
- \cL_Y^{\rm gen} = \bar Q_L \left[ \frac{\sqrt{2}}{v} M_{d} \Phi_v + Z_{d} \Phi_H \right]D_R 
+ \bar Q_L \left[ \frac{\sqrt{2}}{v} M_{u} \Phi_v^c + Z_{u} \Phi_H^c \right] U_R  +{\rm h.c.}~,
\label{eq:generalY2}
\end{eqnarray}
with the $3\times 3$ matrices $Z_{d,u}$ given by
\be
Z_d =  c_\beta X_{d2} - s_\beta X_{d1}~, \qquad 
Z_u =  c_\beta X_{u2} - s_\beta X_{u1}~.
\label{eq:Zdef}
\ee 
It is then straightforward to check that  $Z_d \propto M_d$ and $Z_u \propto M_u$
for all cases in Eqs.~(\ref{eq:NFCII}), (\ref{eq:NFCI}), and (\ref{eq:MFV0}). 
This implies that quark mass terms and couplings to the neutral Higgs fields 
can be diagonalized simultaneously, resulting in the absence of tree-level FCNCs.
Note also that the two NFC structures in Eq.~(\ref{eq:NFCII}) 
and (\ref{eq:NFCI}) correspond to specific limits for the  $c_i$ coefficients 
of the linear MFV structure in~Eq.~(\ref{eq:MFV0}), as recently pointed 
out in Ref.~\cite{Pich:2009sp}.
Less trivial is to understand how these Yukawa 
interactions get modified after the inclusion of quantum corrections. 
This is the subject of the next two sections.

\section{The problems of NFC beyond the lowest order}
\label{sec:NFC}

As discussed in the previous section, 
the NFC hypothesis can be enforced by means of appropriate 
{\em flavour-blind} symmetries. As we will show in the following, these symmetries alone
are not sufficient to protect the effective Yukawa interaction beyond the lowest order. 

\subsection{\boldmath   Breaking of $U(1)_{\rm PQ}$ beyond the tree-level}

We consider first the case where 
the type-II structure in Eq.~(\ref{eq:NFCII}) is enforced 
by means of the $U(1)_{PQ}$ symmetry. In this case  
the structure is not stable since this continuous symmetry must be explicitly 
broken in other sectors of the theory in order to avoid a massless 
pseudoscalar Higgs field.
The $U(1)_{PQ}$ breaking will then induce 
non-vanishing $X_{u1}$ and $X_{d2}$ beyond the tree-level.
If the underlying theory contains additional sources of flavour 
symmetry breaking beside the quark Yukawa couplings (i.e.~if the 
theory is not compatible with the MFV hypothesis), 
the loop-induced couplings $X_{u1}$ and $X_{d2}$ may lead to very large 
FCNCs. This is for instance what happens in the {minimal supersymmetric
extension of the SM (MSSM)} with generic
soft-breaking terms~\cite{Isidori:2002qe,Chankowski:2000ng}.\footnote{~In the absence
of an explicit breaking, the $U(1)_{PQ}$ symmetry would be spontaneously broken by the vev of $H_1$,
hence the theory would contain a Goldstone boson. This problem can be avoided, and
$U(1)_{PQ}$ does not need to be explicitly broken, if the vev
of $H_1$ is zero (see e.g.~Ref.~\cite{Dobrescu:2010rh}).
However, also in this case the smallness of FCNCs is not guaranteed beyond
the tree level (in the absence of MFV) because of the argument presented in Sect.~\ref{sect:Discrete}.}

To quantify the amount of fine-tuning in this scenario in the presence of 
$U(1)_{\rm PQ}$ breaking but not imposing
 MFV, we consider in detail the case of the down-type Yukawa coupling. 
After the breaking of the NFC relation, 
$X_{d1}$ and $X_{d2}$ can be decomposed as
\be
X_{d1} = Y_d~, \qquad X_{d2} = \epsilon_{d} \Delta_{d}~,  
\ee
where $\Delta_d$ is a generic $3\times 3$ flavour-breaking
matrix, with $\cO(1)$ entries, and $\epsilon_{d}$ is a real 
parameter controlling the size of the $U(1)_{PQ}$ breaking.
Since the breaking of $U(1)_{PQ}$ is generated only 
beyond the tree-level, we can assume $\epsilon_{d} \ll 1$.
In the basis where $Y_d$ is diagonal the  down-type mass matrix in 
(\ref{eq:MdMu}) assumes the form
\be
(M_d)_{ij} = \frac{v_1}{\sqrt{2}} 
\left[  (Y^{\rm eff}_d)_{ii}  \delta_{ij} + \epsilon_{d} t_\beta (\widetilde \Delta_{d})_{ij} 
\right]~,
\ee
where $(Y^{\rm eff}_d)_{ii}=(Y_d)_{ii}+\epsilon_{d} t_\beta (\Delta_{d})_{ii}$
and $\widetilde \Delta_{d}$ is the off-diagonal part of $\Delta_{d}$.
We can then proceed with a perturbative diagonalization of $M_d$ to 
first order in $\epsilon_d$. This is obtained via the rotations 
\bea
Q^i_L &\to&  \left[ \delta_{ij} + \epsilon_{d} t_\beta \frac{ (\widetilde \Delta_d)_{ij} 
(Y^{\rm eff}_d)_{jj} + (\widetilde \Delta_d)^*_{ji} (Y^{\rm eff}_d)_{ii} }{
(Y^{\rm eff}_d)^2_{jj}  - (Y^{\rm eff}_d)^2_{ii} } \right] Q^j_L~, \\
D^i_R &\to&  \left[ \delta_{ij} + \epsilon_{d} t_\beta \frac{ (\widetilde \Delta_d)^*_{ji} 
(Y^{\rm eff}_d)_{jj} + (\widetilde \Delta_d)_{ij} (Y^{\rm eff}_d)_{ii} }{
(Y^{\rm eff}_d)^2_{jj}  - (Y^{\rm eff}_d)^2_{ii} } \right] D^j_R~. 
\eea
In the basis where  $M_d$ is diagonal 
the effective coupling $Z_d$ defined in Eq.~(\ref{eq:Zdef})
assumes the form
\bea
(Z_d)_{ij} = (Z^{\rm diag}_d)_{ii} \delta_{ij} 
+ \frac{\epsilon_{d}}{c_\beta}  (\widetilde \Delta_d)_{ij}~,
\qquad (Z^{\rm diag}_d)_{ii} = -s_\beta (Y_d)_{ii} + c_\beta \epsilon_{d} (\Delta_d)_{ii}~,
\eea
which implies the following FCNC coupling:
\be
\cL^{\rm FCNC}_{\epsilon } = - \frac{\epsilon_d}{c_\beta} 
(\widetilde \Delta_{d})_{ij}~
\bar d^i_L  d^j_R~\frac{S_2 + i S_3}{\sqrt{2}}~+~{\rm h.c.},
\label{eq:LFCNC1}
\ee
where $S_{2,3}$ are the neutral components of the Higgs doublet
with vanishing vev (see Appendix).
For $(\widetilde \Delta_{d})_{ij} =\cO(1)$ and $\epsilon_d = \cO(10^{-2})$,
as expected by a typical loop suppression, this effective coupling 
is well above the experimental bounds on FCNCs. In particular,
it largely exceeds the bounds from CP-violation in 
$K^0$--$\bar K^0$ mixing.

\subsection{\boldmath The $\epsK$ bound on generic scalar FCNCs}
To evaluate the impact of the FCNC coupling in Eq.~(\ref{eq:LFCNC1}),
we consider the simplifying case where 
the mass mixing between $\Phi_H$ and $\Phi_v$,
and possible CP-violating terms in the Higgs potential
can be neglected (the so-called decoupling limit, that 
is naturally realized for $t_\beta \gg 1$, see Appendix).
In this limit the neutral components of $\Phi_H$ are 
the CP-even and CP-odd mass-eigenstates $H^0$ and $A^0$, 
with degenerate mass $M_H$. 

Integrating out the heavy Higgs fields at the tree-level 
leads to the following $\Delta S=2$ effective Hamiltonian 
\be
\cH^{\rm |\Delta S|=2}_{\epsilon } = - \frac{ \epsilon_d^2  }{ c^2_\beta M_H^2} 
(\widetilde \Delta_{d})_{21} (\widetilde \Delta_{d})^*_{12} 
(\bar s_L  d_R)(\bar s_R  d_L)~+~{\rm h.c.}~.
\label{eq:HDF2gen}
\ee
Taking into account the large QCD corrections in the evolution from 
$\mu\sim M_H$ down to a scale $\mu_K\sim 2$~GeV, 
this effective Hamiltonian
 implies a potentially sizable non-standard contribution to $\epsK$
(see Section~\ref{sec:DF2}). 
Imposing the condition $| \epsK^{\text{NP}}| < 0.2 |\epsK^{\rm exp}|$, to be in agreement 
with experiment, leads to the bound
\be
| \epsilon_d | \times \left|\text{Im}[(\widetilde \Delta_{d})^*_{21} (\widetilde \Delta_{d})_{12}] 
\right|^{1/2}  \lsim  3 \times 10^{-7}  \times  \frac{ c_\beta M_H}{100~{\rm GeV}}~.
\label{eq:epsdbound}
\ee
This result illustrates the large amount of fine-tuning needed on $\epsilon_d$
if the new flavour-breaking matrix $\Delta_{d}$ has entries of $\cO(1)$: a loop suppression of 
$\cO(10^{-2})$ on  $\epsilon_d$ is not enough to avoid a huge contribution 
to $\epsK$, if $M_H \lsim 1$~TeV. 
In other words, we {\em cannot avoid an efficient protection 
of the flavour structure}, if we want to avoid too large FCNCs.
One could of course suppress FCNCs choosing a very large value for
$M_H$, but this would introduce a fine-tuning problem in the Higgs sector.

The MFV hypothesis is not the only allowed possibility to reach a
sufficiently small breaking of the $SU(3)_q^3$ flavour symmetry. 
For instance, in models with warped space-time geometry~\cite{Agashe:2004cp}
or, equivalently, models with partial compositeness~\cite{Contino:2006nn},
or hierarchical fermion wave functions~\cite{Davidson:2007si}, we expect
\be
|(\widetilde \Delta_{d})^*_{ij} (\widetilde \Delta_{d})_{ji} |_{\rm RS-GIM} = 
\cO(1) \times [(Y_d)_{ii} (Y_d)_{jj}] = \cO(1) \times \frac{ 2 m_{d_i} m_{d_j} }{c_\beta^2 v^2}~,
\label{eq:RSGIM}
\ee  
where the quark masses have to be evaluated at the scale $\mu\sim M_H$.
Using the above relation to set a bound on 
$\text{Im}[(\widetilde \Delta_{d})^*_{21} (\widetilde \Delta_{d})_{12}]$, with the 
unknown $\cO(1)$ coefficients fixed to 1, leads to 
\be
| \epsilon_d |_{\rm RS-GIM}  \lsim   4\times 10^{-3}  \times  \frac{c^2_\beta M_H}{100~{\rm GeV}}~.
\label{eq:RSGIMbound}
\ee  
In this case a $\cO(10^{-2})$ suppression on $\epsilon_d$ and a not too light $M_H$ 
could be sufficient to avoid the $\epsK$ bound, but only if $t_\beta =\cO(1)$.
If $t_\beta$ is large, then also in this case a non-negligible amount of 
fine-tuning is needed.  Moreover, once the  $\epsK$ bound is enforced, 
non-standard effects in $\Delta B=2$ amplitudes are naturally 
suppressed, { unless some amount of fine-tuning 
on the $\cO(1)$ coefficients (the five-dimensional Yukawa couplings) 
is introduced~\cite{Blanke:2008zb}.}

As we will show in Section~\ref{sec:MFV}, the picture is quite different in the MFV case.
Within the MFV framework 
large values of  $t_\beta$ cannot be excluded, and the most interesting phenomenology is expected in
the $B_{s,d}$-meson systems.

\subsection{\boldmath Discrete symmetries and higher-dimensional Yukawa-type interactions}
\label{sect:Discrete}
To derive the effective FCNC coupling in Eq.~(\ref{eq:LFCNC1})
we assumed that the type-II NFC structure of the dimension-four
Yukawa couplings is violated at the quantum level as a consequence 
of the breaking of the $U(1)_{\rm PQ}$ symmetry. 
If the symmetry used to enforce the NFC structure is a discrete one,
this is not necessarily true: we can conceive 
a NFC scenario where $U(1)_{PQ}$ is explicitly broken, while  the $Z_2$
symmetry $H_1 \to - H_1$, $D_R \to - D_R$ is exact. In this case 
$X_{d2}$ and  $X_{u1}$ are strictly zero. However, also this condition 
is not sufficient to protect FCNCs if the theory has additional degrees 
of freedom at the TeV scale, as expected by a natural stabilization 
of the mechanism of electroweak symmetry breaking. 

Integrating out the heavy fields at the TeV scale in a $Z_2$ invariant 
framework generates higher-dimensional operators of the type
\bea
\Delta \cL_Y &=&  \frac{c_1}{\Lambda^2} \bar Q_L X^{(6)}_{u1} U_R H_2 |H_1|^2
+ \frac{c_2}{\Lambda^2} \bar Q_L X^{(6)}_{u2} U_R H_2 |H_2|^2 \nn  \\
&& + \frac{c_3}{\Lambda^2} \bar Q_L X^{(6)}_{d1} D_R H_1 |H_1|^2
+ \frac{c_4}{\Lambda^2} \bar Q_L X^{(6)}_{d2} D_R H_1 |H_2|^2~,
\eea
with $c_i = \cO(1)$ and $\Lambda =\cO(1~{\rm TeV})$. These operators are
$Z_2$ invariant. However, after the Higgs fields get a vev, they 
break the proportionality relation between quark mass terms and effective 
interaction with the neutral scalars, even in the case of a single 
Higgs doublet~\cite{Giudice:2008uua,Agashe:2009di,Azatov:2009na}. As a result, after the mass 
diagonalization, $\Delta \cL_Y$ leads to effective FCNC couplings of 
the type  in Eq.~(\ref{eq:LFCNC1}), where also the physical 
component of the $\Phi_v$ doublet appears. 

In this context the role of the Peccei-Quinn symmetry breaking term $\epsilon_{d}$ 
is replaced by a parameter of order  $v^2/\Lambda^2$. From the bound in Eq.~(\ref{eq:epsdbound}), 
it is clear that for $\Lambda =\cO(1~{\rm TeV})$ this suppression is not
sufficient to be in agreement with data, unless the flavour structure 
of the $X^{(6)}_{i}$ is sufficiently protected.

\section{\boldmath Stability of MFV beyond the lowest order}
\label{sec:MFV}
\subsection{\boldmath  General considerations}
The general structure implied by the MFV hypothesis for the renormalizable 
Yukawa couplings defined in (\ref{eq:generalcouplings}) is 
\bea
X_{d1} &=& P_{d1}(Y_u Y_u^\dagger, Y_d Y_d^\dagger) \times Y_d~,  \\
X_{d2} &=& P_{d2}(Y_u Y_u^\dagger, Y_d Y_d^\dagger) \times Y_d~, \\
X_{u1} &=& P_{u1}(Y_u Y_u^\dagger, Y_d Y_d^\dagger) \times Y_u~, \\
X_{u2} &=& P_{21}(Y_u Y_u^\dagger, Y_d Y_d^\dagger) \times Y_u~, 
\eea
where $P_{i}(Y_u Y_u^\dagger, Y_d Y_d^\dagger)$ are generic polynomials of the 
two basic left-handed spurions
\be
Y_u Y_u^\dagger,~Y_d Y_d^\dagger \sim 
(8, 1,1)_{{\rm SU}(3)^3_q}\oplus(1, 1,1)_{{\rm SU}(3)^3_q}~.
\ee
Since we are free to 
re-define the two basic spurions $Y_u$ and $Y_d$, without loss of generality 
we can define them to be the flavour structures appearing in $X_{d1}$ and $X_{u2}$. Then expanding 
the remaining non-trivial polynomials in powers of $Y_u^\dagger Y_u$ and $Y_d^\dagger Y_d$ 
we get
\bea
X_{d1} &=& Y_d~,  \nn \\
X_{d2} &=& \epsilon_{0} Y_d + \epsilon_{1} Y_d  Y_d^\dagger Y_d                    
+  \epsilon_{2}  Y_u Y_u^\dagger Y_d + \ldots~,  \nn \\
X_{u1} &=& \epsilon^\prime_{0} Y_u + \epsilon^\prime_{1}  Y_u Y_u^\dagger Y_u 
+  \epsilon^\prime_{2}  Y_d Y_d^\dagger Y_u + \ldots~, \nn \\
X_{u2} &=& Y_u~.
\label{eq:XMFVgen} 
\eea
This structure, which has been considered first in full generality in
Ref.~\cite{D'Ambrosio:2002ex}, is renormalization group (RG)
invariant. It is the most general form compatible
with the breaking of the flavour group  $SU(3)_q^3$ by the two 
spurions in Eq.~(\ref{eq:MFVspur}). Quantum corrections can change 
the values of the $\epsilon_i$ at different energy  scales, but they 
cannot modify this functional form.\footnote{~The Yukawa expansion in Eq.~(\ref{eq:XMFVgen})
is very similar to the expansion of the soft-breaking terms in the MSSM, for which 
explicit studies of RG equations in the MFV framework have been discussed in 
Ref.~\cite{Paradisi:2008qh,Colangelo:2008qp}.}
We stress that this functional form is respected only if the 
full theory, including possible high-energy degrees of freedom, 
respects the MFV principle. On the contrary, if we start 
from the linear structure in Eq.~(\ref{eq:MFV0}) but we do not assume 
the MFV principle, we may end up with the problems discussed in the previous section
when going beyond the tree-level.

In principle, the series in Eq.~(\ref{eq:XMFVgen}) contain 
an infinite number of terms. However, barring fine-tuned 
scenarios where $Y_d$ and $Y_u$ have a structure substantially
different than what determined in the one-Higgs case, 
we can still perform the usual MFV expansion in powers of
suppressed off-diagonal CKM elements. More explicitly,  
with an appropriate rotation of the quark fields
we can always choose a basis such that 
\bea
Y_d & \stackrel{\rm d-basis}{\longrightarrow}   & {\rm diag}(\hat y_d,\hat y_s,\hat y_b) \equiv \hat\lambda_d~, \nn  \\
Y_u & \stackrel{\rm d-basis}{\longrightarrow}  & \hatV^\dagger \times {\rm diag}(\hat y_u,\hat y_c,\hat y_t)
\equiv \hatV^\dagger \hat\lambda_u~,
\label{eq:dbasis}
\eea
where the hat over $V$ and the Yukawa eigenvalues distinguish them from
the ``standard'' values obtained in the $\epsilon_i^{(\prime)} =0$ limit:
\bea
\hat \lambda_d &\stackrel{\epsilon^{(\prime)}_i=0}{\longrightarrow}& 
\lambda_d = {\rm diag}( y_d, y_s, y_b)~, \qquad y_{d_i} = \sqrt{2} m_{d_i}/v_{1}~, \\
\hat \lambda_u &\stackrel{\epsilon^{(\prime)}_i=0}{\longrightarrow}& 
\lambda_u = {\rm diag}(y_u, y_c, y_t)~, \qquad y_{u_i} = \sqrt{2} m_{u_i}/v_{2}~, \\
\hat V &\stackrel{\epsilon^{(\prime)}_i=0}{\longrightarrow}& V~[={\rm CKM~matrix}]~. 
\eea
While we cannot fully determine the $\hat V$ and $\hat \lambda_{d,u}$ without
knowing the values of the $\epsilon^{(\prime)}_i$, the smallness of the 
off-diagonal elements of $\hat V$ and of the Yukawa eigenvalues of the
first two generations is parametrically stable, even for 
values of $\epsilon^{(\prime)}_i t_\beta = \cO(1)$. As a result, 
the only large entries in the series~(\ref{eq:XMFVgen}) are 
those involving flavour-diagonal entries of the third generation
and we are left with only two relevant basic spurions
in the basis (\ref{eq:dbasis}). Adopting the notation of
Ref.~\cite{D'Ambrosio:2002ex} we define them as
\be
\Delta = \frac{1}{\hat y^2_b}  Y_d Y_d^\dagger \approx
{\rm diag}(0,0,1)~, \qquad (\hatFC)_{ij} = \left\{ \ba{ll} 
( Y_u Y_u^\dagger )_{ij} 
\approx \hat y_t^2  \hatV^*_{3i} \hatV_{3j}~ &\quad i \not= j~, \\ 0 &\quad i = j~. 
\ea \right.
\label{eq:lhat}
\ee
Expanding to first  non-trivial order in these two spurions
we get~\cite{D'Ambrosio:2002ex}
\beqa
X_{d2} &=& \left( \epsilon_0 + \epsilon_1  \Delta + \epsilon_2
 \hatFC + \epsilon_3 \hatFC \Delta
 +\epsilon_4 \Delta \hatFC \right) {\hat\lambda}_d~, \label{eq:LYPQ_D} \\
X_{u1}  &=& \left( \epsilon'_0 + \epsilon'_1  \Delta + \epsilon'_2
 \hatFC + \epsilon'_3 \hatFC \Delta 
 +\epsilon'_4 \Delta \hatFC \right) \hatV^\dagger {\hat\lambda}_u~. \label{eq:LYPQ_U}
\eeqa
We stress that this form is not a simple linear expansion in the Yukawa couplings,
rather an expansion in the small terms associated to off-diagonal CKM matrix elements
and light quark masses.
The resummation to all orders of high-powers of $\hat y^2_t$ or $\hat y^2_b$, 
whose importance has been stressed in~\cite{Kagan:2009bn,Feldmann:2008ja}, 
is implicitly taken into account by a redefinition of the $\epsilon_i^{(\prime)}$
parameters. 
Throughout this paper we also assume the $\epsilon_i^{(i)}$ are small
($\epsilon_i^{(i)} <1$)
as resulting from an approximate $U(1)_{\rm PQ}$ symmetry. Beside its phenomenological
interest, this assumption allows us to unambiguously define $t_\beta$ starting 
from the $\epsilon_i^{(i)} \to 0$ limit (see Appendix).

The diagonalization of the quark mass matrices keeping 
the $\epsilon_i t_\beta$ terms to all orders
(assuming real $\epsilon_i$ and neglecting $\epsilon_i^\prime/t_\beta$)
has been presented in Ref.~\cite{D'Ambrosio:2002ex}
(see also~\cite{Babu:1999hn,Isidori:2001fv,Buras:2002vd,Dedes:2002er}) 
and will not be repeated here. The main results
can be summarized as follows:
\begin{itemize}
\item{}
The relation between $\lambda_d$ and $\hat \lambda_d$ is
\be
\lambda_d = \left[1+  (\epsilon_0 +\epsilon_1\Delta) t_\beta \right] {\hat\lambda}_d~,
\ee
while the up-type mass matrix remains unaffected ($\hat \lambda_u = \lambda_u$)
in the limit were we neglect 
$\cO(\epsilon^\prime_i/t_\beta)$ terms.  The physical CKM matrix
coincides with $\hatV$ but for the $V_{i3}$ and $V_{3i}$
entries ($i\not=3$) for which 
\beq
\frac{ \hatV_{i3} }{V_{i3} } =
\frac{ \hatV_{3i} }{V_{3i} } = 1+r_V~,  \qquad 
r_V \equiv \frac{(\epsilon_2+\epsilon_3) t_\beta}{1 +
      (\epsilon_0+\epsilon_1-\epsilon_2-\epsilon_3) t_\beta}~.
\eeq
\item{}
The diagonalization of the mass terms does not eliminate
scalar FCNC interactions. In the case of down-type
quarks, the effective FCNC coupling surviving after the
diagonalization can be written as 
\beq
\cL_{\rm MFV}^{\rm FCNC} ~=~ - \frac{1}{s_\beta}
~ {\bar d}^i_L \left[
\left( a_0 V^\dagger \lambda_u^2 V + a_1 V^\dagger \lambda_u^2 V  \Delta
     + a_2  \Delta V^\dagger \lambda_u^2 V \right) \lambda_d \right]_{ij}
 d^j_R~\frac{S_2 + i S_3}{\sqrt{2}} {\rm ~+~h.c.},
\label{eq:LH_FCNC}
\eeq
where $S_{2,3}$ are the neutral components of the Higgs doublet
with no vev ($\Phi_H$, see also Appendix), and 
\beqa
a_0 &=& \frac{  \epsilon_2 t_\beta (1+r_V)^2 }{ y_t^2
       \left[ 1+ \epsilon_0 t_\beta\right]^2}~, \qquad\qquad
a_1 +a_0 = \frac{ r_V }{y_t^2 \left[ 1+ (\epsilon_0+\epsilon_1)t_\beta
\right]}~, \label{eq:ai} \nn \\
a_2 -a_1 &=& \frac{(\epsilon_4-\epsilon_3 )t_\beta}{y_t^2
\left[1+\epsilon_0 t_\beta\right]
       \left[1 + (\epsilon_0+\epsilon_1-\epsilon_2-\epsilon_3)t_\beta
\right]}~.
\eeqa
\end{itemize}
In principle, a FCNC coupling with the $\Phi_H$ doublet survives also in the 
up sector.  However, this effect is less interesting 
since in this case the $a'_i$  coefficients (defined
in analogy with the $a_i$) turn out to be $\cO(\epsilon'_i)$
and not of $\cO(\epsilon_i t_\beta)$ as in~(\ref{eq:ai}).

As can be noted, $\cL_{\rm MFV}^{\rm FCNC}$ exhibits the typical 
MFV structure of FCNCs, where all the non-vanishing effects are 
driven by the large top-quark Yukawa coupling. This structure 
implies a strong suppression of FCNCs because of the smallness
of the CKM elements $|V_{ts}|$ and $|V_{td}|$. As a result, 
the $a_i$ can be of $\cO(1)$ even for $t_\beta \gg 1$ and $M_H<1$~TeV
(detailed phenomenological bounds are presented in Section~\ref{sec:pheno}).
However, the presence of the $\Delta$ spurion, which reflects the 
possibility of large bottom Yukawa coupling, implies a
possible $\cO(1)$ breaking of the correlation 
between FCNCs in the $K$  and in the $B_{s,d}$ systems 
that holds in the SM and in MFV for $t_\beta =\cO(1)$.

\subsection{\boldmath Introducing flavour-blind phases}
\label{sect:fbphases}

So far, following Ref.~\cite{D'Ambrosio:2002ex},
we have assumed that the $\epsilon_i$ are real. This assumption is 
justified by the strong bounds on flavour-conserving CPV phases 
implied by the electric dipole moments\footnote{As the recent analysis \cite{Buras:2010zm} shows, in 
the $\text{2HDM}_{\overline{\text{MFV}}}$ the electric dipole moments, although very much enhanced over the SM values, are still compatible with the experimental bounds.}. However, this phenomenological 
problem could have a dynamical explanation and it is worth to investigate
also the case of complex $\epsilon_i$, namely the case where we allow 
generic CP-violating flavour-blind phases in the Higgs sector.
 Namely, we consider the generic framework where the Yukawa matrices 
are the only sources of breaking of the $SU(3)_q^3$ flavour group,
but they are not the only allowed sources of CP-violation 
(a general discussion about this framework 
can be found in Ref.~\cite{Kagan:2009bn,Mercolli:2009ns,Paradisi:2009ey}).

As far as Higgs-mediated FCNCs are concerned, this amounts only to 
consider the $a_i$ in Eq.~(\ref{eq:LH_FCNC}) as complex parameters. 
Integrating out the neutral Higgs fields leads to 
tree-level contributions to scalar FCNC operators. 
Keeping complex $a_i$,
and working in the decoupling limit for the heavy Higgs doublet
(see Appendix), the leading $\Delta F=1$ and  $\Delta F=2$ 
Hamiltonians thus generated are
\bea
 \cH_{\rm MFV}^{|\Delta B|=1} &=&
 -\frac{a^*_0+a^*_1}{M_H^2}~ y_\ell y_b y_t^2 V^*_{tb} V_{tq}~
 ({\bar b}_R q_L)({\bar \ell}_L \ell_R) {\rm ~+~h.c.} \quad (q=d,s)~, 
\label{eq:HeffBll} \\
 \cH_{\rm MFV}^{|\Delta S|=1} &=&
 -\frac{a^*_0}{M_H^2}~ y_\ell y_s y_t^2 V^*_{ts} V_{td}~
 ({\bar s}_R d_L)({\bar \ell}_L \ell_R) {\rm ~+~h.c.}~,  \\
 \cH_{\rm MFV}^{|\Delta B|=2} &=&
 - \frac{(a^*_0+a^*_1)(a_0+a_2)}{  M_H^2} ~y_b y_q [y_t^2 V^*_{tb} V_{tq} ]^2~
 ({\bar b}_R q_L) ({\bar b}_L q_R)  {\rm ~+~h.c.} \quad (q=d,s)~, \quad \\
 \cH_{\rm MFV}^{|\Delta S|=2} &=&
 - \frac{|a_0|^2}{  M_H^2} ~y_s y_d [y_t^2 V^*_{ts} V_{td} ]^2~
 ({\bar s}_R d_L) ({\bar s}_L d_R)  {\rm ~+~h.c.}~.
\eea
A detailed analysis of the phenomenological impact of these 
effective Hamiltonians is postponed to Section~\ref{sec:pheno}. 
We anticipate here two key properties that can be directly 
deduced by looking at their flavour- and CP-violating structure:
\begin{enumerate}
\item[I.]
The impact in $K^0$--$\bar{K}^0$,  $B^0_{d}$--$\bar{B}^0_{d}$
and $B^0_{s}$--$\bar{B}^0_{s}$ mixing amplitudes scales, relative to the SM,
with $m_sm_d$, $m_b m_d$ and $m_b m_s$, respectively.
This fact opens the possibility of sizable non-standard 
contributions to the $B_s$ system  without serious constraints 
from  $K^0$--$\bar{K}^0$ and  $B^0_{d}$--$\bar{B}^0_{d}$ mixing.
\item[II.]
While the possible flavour-blind phases do not contribute to the 
$\Delta S=2$ effective Hamiltonian, they could have an impact in the 
$\Delta B=2$ case, offering the possibility to solve the 
recent experimental anomalies related to the $B_s$ mixing phase. 
However, this happens only if the $\epsilon_{3,4}$ 
terms in the expansion (\ref{eq:LYPQ_D}) are at least as large 
as the other $\epsilon_i$. This implies non-trivial underlying dynamical 
models, where effective operators with high powers of Yukawa insertions 
are not strongly suppressed (contrary to what happens, for instance, 
in the MSSM).
\end{enumerate}
We emphasize that the CKM and quark-mass pattern of $\Delta F=2$ amplitudes 
outlined above (point I.) is characteristic of the MFV pattern in its general 
formulation~\cite{D'Ambrosio:2002ex}. In particular,  it differs from the 
so-called constrained MFV framework~\cite{Buras:2000dm}, where 
scalar amplitudes are negligible and the three down-type $\Delta F=2$ amplitudes have a 
universal scaling relative to the SM (i.e.~the scaling is dictated only
by the CKM factors). It differs also from the RS-GIM 
(or hierarchical wave-function) framework~\cite{Agashe:2004cp,Contino:2006nn,Davidson:2007si}, 
where we expect similar quark-mass 
suppression terms, but we do not expect also the CKM factors: as shown
in Section~\ref{sect:phenoDF2gen},  the RS-GIM scaling implies large effects 
in $\epsK$, relative to the SM, with minor corrections in the $B_{s,d}$ systems.

Within the general MFV framework this pattern is not necessarily 
a signal of Higgs-mediated effects,
{but it is a clear signal of  
non-Hermitian scalar operators of the type
\be
 (\bar D_R Y_d^\dagger Y_d^{\phantom{\dagger}} Y_d^\dagger Y_u^{\phantom{\dagger}} Y_u^\dagger Q_L)
 (\bar Q_L Y_u^{\phantom{\dagger}} Y_u^\dagger Y_d^{\phantom{\dagger}} D_R)~, \quad 
 (\bar D_R Y_d^\dagger Y_u^{\phantom{\dagger}} Y_u^\dagger Q_L)
 (\bar Q_L  Y_d^{\phantom{\dagger}} Y_d^\dagger Y_u^{\phantom{\dagger}} Y_u^\dagger 
 Y_d^{\phantom{\dagger}} D_R)~, 
\ee
which should appear with a sizable coefficient and 
with large (and non identical) CP-violating phases.}

\section{\boldmath  Comparison with models in the literature}
\label{sec:comparison}

In this section we briefly comment about the flavour structure of the 
models recently proposed in~\cite{Joshipura:2007cs,Pich:2009sp,Botella:2009pq}, 
discussing which of them are compatible 
with the general MFV structure, as described in Section~\ref{sec:Yukawa} and~\ref{sec:MFV}.
To this purpose, it is first useful to clarify 
differences and similarities with respect to Ref.~\cite{D'Ambrosio:2002ex}
of other implementations of the MFV concept present in the literature 
and, more generally, of other constructions 
where the strength of FCNCs is related to the structure of the CKM matrix:
\begin{itemize}
\item{}
The so-called {\em Constrained MFV}, 
originally proposed in~\cite{Buras:2000dm}, is based on the following two assumptions: 
1) no new effective FCNC dimension-six operators beyond those already present in the SM;
2) the flavour and CPV structure of these operators is dictated by the CKM factors
$V_{3i}^*V_{3j}$ (for $d_i \to d_j$ FCNC transitions), i.e.~it is aligned in flavour
space with the SM short-distance contribution. This proposal gives rise to a very 
predictive framework that coincides, in practice, with Ref.~\cite{D'Ambrosio:2002ex} 
in the limiting case of a single light Higgs field.
{However, by construction, the CMFV hypothesis cannot be applied to 
multi-Higgs models with  large $\tan\beta$ where new operators, 
the scalar (left-right) operators, 
can be important. }
\item{}
The so-called {\em General MFV}, proposed in~\cite{Kagan:2009bn}, is based on 
the same symmetry and symmetry-breaking pattern proposed in~\cite{D'Ambrosio:2002ex},
but for the decoupling of CP and flavour symmetries.  In Ref.~\cite{Kagan:2009bn}
particular emphasis is put on the fact that the expansion 
in powers of the top and bottom Yukawa couplings should not be truncated. 
As already discussed in Section~\ref{sec:MFV}, the non-linear structure
associated to the third generation does not give rise to observable differences 
with respect to the formulation of~\cite{D'Ambrosio:2002ex}, which is based 
only on the expansion in off-diagonal CKM elements $|V_{3i}|$
and light quark mass ratios $m_{q_i}/m_{q_3}$ ($i\not=3$). 
The only important difference between \cite{Kagan:2009bn} and~\cite{D'Ambrosio:2002ex} 
is related to the possible introduction of flavour-blind phases, namely to the 
possible decoupling between the breaking of CP and flavour symmetries.
\item{}
The so-called {\em BGL models}, proposed in~\cite{Branco:1996bq}, is a class of
two-Higgs doublet models where the strength of FCNCs in the up- or down-type sector 
is unambiguously related to the off-diagonal elements of the CKM matrix. 
While all the six BGL models are interesting, only one of them is
compatible with the MFV principle. As can be deduced by looking at Eq.~(\ref{eq:lhat}),
only the BGL model where  $d_i \to d_j$ FCNC transitions are proportional to $V_{3i}^*V_{3j}$
is an explicit example of MFV.
\end{itemize}

\noindent 
We are now ready to comment about the explicit models with more than one-Higgs 
doublet considered in  Ref.~\cite{Joshipura:2007cs,Pich:2009sp,Botella:2009pq}:
\begin{itemize}
\item{}
As already pointed out in Section~\ref{sec:Yukawa}, the model of Ref.~\cite{Pich:2009sp}, 
denoted ``Yukawa alignment model'', is a limiting case of the general MFV construction, 
where the higher-order powers in $Y_u$ and $Y_d$ are not included. 
As already pointed out, the higher-order terms are naturally generated by
quantum corrections, but they don't spoil the nice virtues of the MFV construction,
once we assume that the MFV hypothesis is respected by the 
high-energy degrees of freedom of the theory.
Being compatible with the MFV hypothesis, 
some phenomenological aspects of the model of Ref.~\cite{Pich:2009sp}
can be deduced as a limiting case from more general MFV analyses 
(see e.g.~\cite{D'Ambrosio:2002ex,Babu:1999hn,Isidori:2001fv,Buras:2002vd,Dedes:2002er}).  
However it must be stressed that, except for Ref.~\cite{Pich:2009sp}, 
previous works have not included the possibility of flavour-blind
CP-violating phases, and have been focused mainly on neutral-Higgs effects. 
\item{}
The first model considered in Ref.~\cite{Botella:2009pq}, namely 
a 2HDM with a non-trivial $U(1)$ symmetry for the third generation 
is an interesting explicit example of MFV. 
More precisely, this framework coincides with the MFV construction 
in the limit $m_{c,u}^2/m_t^2 \to 0$, which is an excellent approximation.
On the other hand, the model with three-Higgs doublet considered
at the end of  Ref.~\cite{Botella:2009pq} is not compatible with 
the MFV principle and, not surprisingly, it leads to potentially
too large effects in $K^0$--$\bar K^0$ mixing.
\item{} 
As far as Ref.~\cite{Joshipura:2007cs} is concerned, the model considered 
in the first paper is, in principle, compatible with MFV. However, 
the related phenomenological analysis is not reliable since the 
authors have assumed a huge $SU(2)_L$ breaking in the Higgs sector,
which can be ruled out by electroweak precision tests. Indeed 
they analyse the phenomenological implications of the 
$\bar d^i_L d^j_R  \bar d^i_L d^j_R$ operator that is forbidden 
in the exact $SU(2)_L$ limit (see Section~\ref{sect:phenogen}). 
The model considered in the second paper 
in Ref.~\cite{Joshipura:2007cs} is manifestly beyond MFV.

\end{itemize}

\section{\boldmath  Phenomenological tests}
\label{sec:pheno}

\subsection{\boldmath  $\Delta F=2$ amplitudes: general discussion}
\label{sec:DF2}

In the general 2HDM 
the $\Delta F=2$ transitions represented by the particle-antiparticle mass differences $\Delta M_K$, $\Delta M_{s,d}$ and the CP-violating observables 
$\epsK$, $S_{\psi K_s}$ and $S_{\psi\phi}$ are governed by the SM box diagrams with up-quarks and $W^\pm$ exchanges, the box diagrams with up-quarks and $H^\pm$ exchanges and the tree-level neutral Higgs $(h^0,H^0,A^0)$ exchanges.

When QCD renormalization group effects are taken into account, 
the following set of low-energy operators at scales $\mu_K\sim 2$ GeV in the 
case of $K^0-\bar K^0$ mixing has to be taken into account:
\be
\begin{array}{rcl}
Q_1^{VLL}&=&(\bar s_L \gamma_\mu d_L)(\bar s_L \gamma^\mu d_L)\,, \\
Q_1^{SLL}&=&(\bar s_R d_L)(\bar s_Rd_L)\,, \\
Q_2^{SLL}&=&(\bar s_R \sigma_{\mu\nu} d_L)(\bar s_R \sigma^{\mu\nu} d_L)\,, 
\end{array}
\qquad 
\begin{array}{rcl}
Q_1^{LR}&=&(\bar s_L \gamma_\mu d_L)(\bar s_R \gamma^\mu d_R)\,, \\
Q_2^{LR}&=&(\bar s_R d_L)(\bar s_Ld_R)\,, \\ & &
\end{array}
\label{eq:operatorsDF2}
\ee
where $\sigma_{\mu\nu}=\frac{1}{2}\left[\gamma_\mu,\gamma_\nu\right]$. In addition, the operators 
$Q_{1,2}^{SRR}$, analogous to $Q_{1,2}^{SLL}$ with the exchange $q_L\rightarrow q_R$, are also present. 
Once the Wilson coefficients of these operators $C_i(\mu_H)$
 are calculated at a high energy scale, where heavy degrees of freedom are integrated out, the formulae in \cite{Buras:2001ra} allow automatically to calculate their values at scales $\mathcal O(\mu_K)$. 

The resulting low energy effective Hamiltonian then reads
\be
\cH_{\rm eff}=\sum_{i,a}C_i^a(\mu_K,K)Q_i^a,
\ee
where $i=1,2$, and  $a=VLL,LR,SLL,SRR$.
The off-diagonal element in $K^0-\bar K^0$ mixing $M_{12}^K$ is then given by
\be
 2 M_K M_{12}^K=\langle \bar K^0|\cH_{\rm eff}|K^0\rangle^*
\ee
and the observables of interest can be directly evaluated. In particular
\bea
\Delta M_K&=&2 \text{Re}\left( M_{12}^K\right)\,,\\
\epsK&=&e^{i\varphi_\epsilon}\frac{\kappa_{\epsilon}}{\sqrt 2\, \Delta M_K}{\text{Im}}\left(M_{12}^K\right)\,.
\eea
 Here $\varphi_\varepsilon = (43.51\pm0.05)^\circ$  takes into account that $\varphi_\varepsilon\ne \pi/4$ and $\kappa_\varepsilon=0.94\pm0.02$ \cite{Buras:2008nn,Buras:2010pz} includes an additional effect from 
long-distance contributions.

Finally for the $B_{s,d}$ systems, one has to evaluate the Wilson coefficients at scales $\mu_B\sim 4.2$ GeV. The computation of the CPV and CP-conserving (CPC) observables is then exactly analogous 
to what we have just discussed for the $K$ system. In particular, denoting with $M_{12}^q$ the off-diagonal 
elements in the $B^0_q-\bar B^0_q$ mixings, one has
\be
\Delta M_q=2\left|M_{12}^q\right|, \qquad (q=d,s)\,.
\ee

Next we define
\begin{equation}
M_{12}^q=\left( M_{12}^q\right)_\text{SM} C_{B_q}e^{ i2\varphi_{B_q}}~, 
\qquad (q=d,s)~,
\label{eq:3.36}
\end{equation}
where 
\bea
(M_{12}^d)_\text{SM} &=& \big| (M_{12}^d)_\text{SM} \big|e^{2i\beta} \,,\qquad\beta\approx 0.38~,
\label{eq:3.38} \\
(M_{12}^s)_\text{SM} &=& \big| (M_{12}^s)_\text{SM} \big|e^{2i\beta_s} \,,\qquad\beta_s\simeq -0.01~,
\label{eq:3.39}
\eea
and the phases $\beta$ and $\beta_s$ are defined through
\begin{equation}
V_{td}=|V_{td}|e^{-i\beta}\quad\textrm{and}\quad V_{ts}=-|V_{ts}|e^{-i\beta_s}\,.
\label{eq:3.40}
\end{equation}
Possible non-standard effects would manifest themselves via 
$C_{B_q}\not=1$,  $\varphi_{B_d}\not=0$, or $\varphi_{B_s}\not=0$. 
Using this notation the physical observables are 
\begin{equation}
\Delta M_q=(\Delta M_q)_\text{SM}C_{B_q}~,
\label{eq:3.41}
\end{equation}
and
\begin{equation}
S_{\psi K_S} = \sin(2\beta+2\varphi_{B_d})\,, \qquad
S_{\psi\phi} =  \sin(2|\beta_s|-2\varphi_{B_s})\,,
\label{eq:3.43}
\end{equation}
with the latter two observables being the coefficients of $\sin(\Delta M_d t)$ and $\sin(\Delta M_s t)$ in the time dependent asymmetries in $B_d^0\to\psi K_S$ and $B_s^0\to\psi\phi$, respectively. Thus in the presence of non-vanishing $\varphi_{B_d}$ and $\varphi_{B_s}$ these two asymmetries do not measure $\beta$ and $\beta_s$ but $(\beta+\varphi_{B_d})$ and $(|\beta_s|-\varphi_{B_s})$, respectively.

While working with Wilson coefficients and operator matrix elements at
low energy scales is a common procedure, it turns out that   
 for phenomenology it 
is more useful to work directly with $C_i(\mu_H)$ and with the 
hadronic matrix elements of  the corresponding operators also evaluated 
at this high scale. The latter matrix elements are given by \cite{Buras:2001ra}
\be
\langle \bar K^0|Q_i^a|K^0\rangle = \frac{2}{3}M_K^2 F_K^2 P_i^a(K)~,
\label{eq:Pidef}
\ee
where the coefficients $P_i^a(K)$, discussed in more detail below, 
collect compactly all RG effects from scales below $\mu_H$ as well as
hadronic matrix elements obtained by lattice methods. 

The off-diagonal element in $K^0-\bar K^0$ mixing $M_{12}^K$ is then given by
\be\label{eq:M12K}
M_{12}^K = \frac{1}{3}M_K F_K^2\sum_{i,a} C_i^{a*}(\mu_H,K) P_i^a(K).
\ee
Similarly we find 
\be\label{eq:M12q}
M_{12}^q = \frac{1}{3}M_{B_q} F_{B_q}^2\sum_{i,a} C_i^{a*}(\mu_H,B_q) P_i^a(B_q).
\ee

The following points should be emphasized
\begin{itemize}
\item The expressions (\ref{eq:M12K}) and (\ref{eq:M12q}) are valid 
for any 2HDM with model dependence entering only the Wilson coefficients $C_i^a(\mu_H)$, 
which generally also depend on the meson system considered. In particular, 
they are valid both within and beyond the MFV framework. In MFV models
CKM factors and Yukawa couplings 
define the flavour dependence of these coefficients, while in 
non-MFV models additional flavour structures are present in $C_i^a(\mu_H)$.
\item The coefficients $P_i^a$ are model independent and include the renormalization group evolution from high scale $\mu_H$ down to low energy $\mathcal O(\mu_K,\mu_B)$. As the physics cannot depend on the renormalization scale $\mu_H$, the 
$P_i^a$ depend also on $\mu_H$ so that the scale dependence present in 
$P_i^a$ is canceled by the one in $C_i^a$. Explicit formulae for $\mu_H$ dependence of $P_i^a$ can be found in\cite{Buras:2001ra}. It should be stressed that here we are talking about logarithmic dependence on $\mu_H$. The power-like dependence (such as $1/M_H^2$, \ldots) is present only in the $C_i^{a}$.
\item The $P_i^a$ depend however on the system considered as the hadronic matrix elements of the operators in (\ref{eq:operatorsDF2}) relevant for $K^0-\bar K^0$ mixing differ from the matrix elements of analogous operators relevant for $B_{s,d}^0-\bar B_{s,d}^0$ systems. Moreover whereas the RG evolution in the latter systems stops at $\mu_B=\mathcal O(M_B)$, in the case of $K^0-\bar K^0$ system it is continued down to $\mu_K\sim 2$ GeV, where the hadronic matrix elements are evaluated by lattice methods.
\end{itemize}

The advantage of formulating everything at the high energy scale will
be evident in the next subsections.

\begin{figure}[t]
\begin{center}
\includegraphics[width=.7\textwidth]{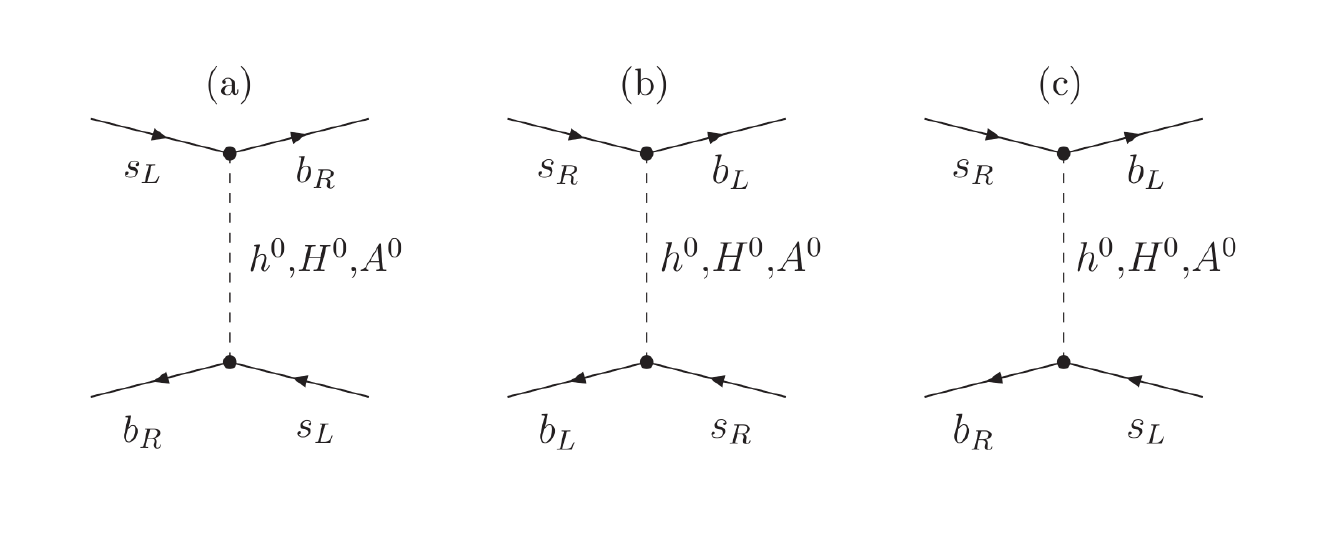}
\vskip -0.8 true cm
\caption{\label{fig:2pg}
Tree-level Higgs-mediated contributions to $\Delta F=2$ amplitudes.}
\end{center}
\end{figure}

\subsection{\boldmath  Neutral-Higgs exchange in $\Delta F=2$ amplitudes: general case}
\label{sect:phenogen}

\subsubsection{Basic formulae} 
Let us next consider the impact of neutral Higgs exchanges on meson-antimeson mixings
generated by the general effective Lagrangian in Eq.~(\ref{eq:LFCNC1}).
The neutral Higgs exchanges contribute to $M_{12}^K$ at tree-level through the diagrams 
in Fig.~\ref{fig:2pg}. Evidently (a), (b) and (c) contribute 
to $C_1^{SLL} (\mu_H)$, $C_1^{SRR} (\mu_H)$ and $C_2^{LR} (\mu_H)$, respectively. 
In the absence of QCD corrections to diagrams in Fig.~\ref{fig:2pg},	
\be
C_1^{LR} (\mu_H)=C_2^{SLL} (\mu_H)=C_2^{SRR} (\mu_H)=0~.	
\ee
Still even in this case the presence of $Q_1^{LR}$, $Q_2^{SLL}$ 
and $Q_2^{SRR}$ enters in  $P_1^{LR}$, $P_2^{SLL}$ and $P_2^{SRR}$ as discussed below and 
 seen explicitly  in the formulae of \cite{Buras:2001ra}.

In our phenomenological analysis (both within MFV and beyond) 
we assume the decoupling limit of the heavy Higgs doublet (see Appendix). 
In this limit the $H^0$ and $A^0$ contributions 
to $C_1^{SLL} (\mu_H)$ and $C_1^{SRR} (\mu_H)$ cancel 
approximately each other and contribute constructively to $C_2^{LR} (\mu_H)$.
This cancellation  is not accidental and can be understood in 
terms of the $SU(2)_L$ structure of the effective operators, since 
only $Q_2^{LR}$ survives in the $SU(2)_L$-invariant limit 
(see~e.g.~\cite{D'Ambrosio:2002ex,Gorbahn:2009pp}).

By construction, the contribution from $h^0$ to all the effective 
operators can be neglected. In this limit, as seen in (\ref{eq:HDF2gen}),
\be
C_2^{LR}(\mu_H,K) = - \frac{ \epsilon_d^2  }{ c^2_\beta M_H^2} 
(\widetilde \Delta_{d})_{21} (\widetilde \Delta_{d})^*_{12} 
\label{eq:CLRK}
\ee
and using (\ref{eq:M12K}) we find
\be\label{eq:M12KHiggs}
M_{12}^K = -\frac{1}{3}M_K F_K^2 P_2^{LR}(K)
\frac{ \epsilon_d^2  }{ c^2_\beta M_H^2} 
(\widetilde \Delta_{d})^*_{21} (\widetilde \Delta_{d})_{12}~. 
\ee
Similarly, we find
\bea
\label{eq:M12dHiggs}
M_{12}^d &=& -\frac{1}{3}M_{B_d} F_{B_d}^2 P_2^{LR}(B_d)
\frac{ \epsilon_d^2  }{ c^2_\beta M_H^2} 
(\widetilde \Delta_{d})^*_{31} (\widetilde \Delta_{d})_{13}~, \\
\label{eq:M12sHiggs}
M_{12}^s &=& -\frac{1}{3}M_{B_s} F_{B_s}^2 P_2^{LR}(B_s)
\frac{ \epsilon_d^2  }{ c^2_\beta M_H^2} 
(\widetilde \Delta_{d})^*_{32} (\widetilde \Delta_{d})_{23}~. 
\eea

\subsubsection{\boldmath  Phenomenology of $K^0$--$\bar{K}^0$ 
and $B^0_{s,d}$--$\bar{B}^0_{s,d}$ systems}
\label{sect:phenoDF2gen}

\begin{table}[t] 
\begin{center}
\begin{tabular}{|l|l||l|l|}
\hline
parameter & value & parameter & value \\
\hline\hline
$F_K$ & $(155.8\pm 1.7) \text{MeV}$	\cite{Laiho:2009eu}	  & $m_s(2\,\text{GeV})$&$0.105\,\text{GeV}$~\cite{Amsler:2008zzb}\\
$F_{B_d}$ & $(192.8 \pm 9.9) \text{MeV}$ \cite{Laiho:2009eu} 	  & $m_d(2\,\text{GeV})$&$0.006\,\text{GeV}$~\cite{Amsler:2008zzb}\\
$F_{B_s}$ & $(238.8 \pm 9.5) \text{MeV}$\cite{Laiho:2009eu}	  & $|V_{ts}|$&$0.040\pm 0.003$~\cite{Bona:2007vi}\\
$\hat B_K$ & $0.725 \pm 0.026$ 	\cite{Laiho:2009eu}		  & $|V_{tb}|$&$ 1\pm 0.06$~\cite{Bona:2007vi}\\
$\hat B_{B_d}$ & $1.26\pm 0.11$ \cite{Laiho:2009eu}		  & $|V_{td}|_{\rm tree}$&$(8.3\pm 0.5)\cdot 10^{-3}$~\cite{Bona:2007vi}\\
$\hat B_{B_s}$ & $1.33\pm 0.06$	\cite{Laiho:2009eu}	          & $|V_{us}|$&$ 0.2255 \pm 0.0019 $\cite{Amsler:2008zzb}\\
$M_{B_s}$&$5.3664$ GeV~\cite{Amsler:2008zzb}                       & $|V_{cb}|$&$ (41.2 \pm 1.1)\times 10^{-3}$\cite{Amsler:2008zzb}\\
$M_{B_d}$&$5.2795$ GeV~\cite{Amsler:2008zzb}                       & $\sin(2\beta)_{\rm tree}$ & $0.734\pm 0.038$~\cite{Bona:2007vi}\\
$M_K$&$0.497614\,\text{GeV}$~\cite{Amsler:2008zzb}		  & $\sin(2\beta_s)$&$0.038\pm 0.003$~\cite{Bona:2007vi}\\ 
$\eta_{cc}$ & $1.51\pm 0.24$\cite{Herrlich:1993yv}		  & $\alpha_s(m_Z)$&$ 0.1184$~\cite{Bethke:2009jm}\\ 
$\eta_{tt}$ & $0.5765\pm 0.0065$\cite{Buras:1990fn}		  & $\Delta M_s$ & $(17.77\pm 0.12)~{\rm ps}^{-1}$ ~\cite{Barberio:2008fa}\\
$\eta_{ct}$ & $0.47\pm 0.04$\cite{Herrlich:1995hh}		  & $\Delta M_d$ & $(0.507\pm 0.005)~ {\rm ps}^{-1}$ ~\cite{Barberio:2008fa}\\
$\eta_{B}$  & $0.551\pm 0.007$\cite{Buras:1990fn,Buchalla:1996ys}  & $\Delta M_K$&$(5.292\pm 0.009)\cdot 10^{-3} ps^{-1}$~\cite{Amsler:2008zzb}\\
$\xi$ & $1.243 \pm 0.028$\cite{Laiho:2009eu}	                  & $\kappa_\varepsilon$& $0.94\pm 0.02$~\cite{Buras:2010pz} \\
$m_c(m_c)$ & $(1.268\pm 0.009) \gev$\cite{Allison:2008xk}	  & $\epsK^{\text{exp}}$&$(2.229\pm 0.01)\cdot 10^{-3}$~\cite{Amsler:2008zzb}\\
$m_t(m_t)$ & $(163.5\pm 1.7) \gev$\cite{:2009ec}		  & $S_{\psi K_S}^{\text{exp}}$& $0.672\pm0.023$~\cite{Barberio:2008fa}\\
$m_b(m_b)$& $ (4.2+0.17-0.07) \gev$~\cite{Amsler:2008zzb}          & & \\ 
\hline
\end{tabular}
\caption{Values of the input parameters used in our analysis. Additionally, the $P_{i}^{a}$ parameters defined in (\ref{eq:Pidef}) are computed using results 
from Ref.~\cite{Allton:1998sm,Becirevic:2001xt}.
The subscript ``tree'' in $|V_{td}|$ and $\sin(2\beta)$ denotes that these inputs are
extracted from data using only tree-level observables~\cite{Bona:2007vi}.} \label{tab:eps}
\end{center}
\end{table}

In the Kaon system the value of $P_2^{LR} (K)$, computed using the formulae 
in \cite{Buras:2001ra}, the input values in Table~\ref{tab:eps} and the
hadronic matrix elements in~\cite{Babich:2006bh}, is 
\be
P_2^{LR} (K) \approx 66 \qquad {\rm for} \qquad  \mu_H=246~{\rm GeV}.
\ee 
This is about two order of magnitude larger 
than the corresponding factor for the SM operator: $P_1^{VLL} (K) \approx 0.42$. 
This difference originates from the strong renormalization group enhancement 
and the chiral enhancement of the scalar operator $Q_2^{LR}$. Consequently, 
even a small  new physics contribution to $C_2^{LR} (\mu_H)$ 
can play an important role in the phenomenology.

Using (\ref{eq:M12KHiggs}) and the input parameters in Table~\ref{tab:eps}
we find then for the contribution of the neutral Higgs exchanges to $\epsK$:
\begin{equation}
\Delta \epsK  = 
-7.7\times 10^{11}\gev^2\frac{\epsilon_d^2 }{ c^2_\beta M_H^2} P_2^{LR} (K)~
 \text{Im}[ (\widetilde \Delta_{d})^*_{21} (\widetilde \Delta_{d})_{12}]~,
\end{equation}
leading for $P_2^{LR} (K) = 66$ to the bound in (\ref{eq:epsdbound}).

The formulae for $M_{12}^q$ relevant for the $B^0_{s,d}$--$\bar{B}^0_{s,d}$ systems
have been given in (\ref{eq:M12dHiggs}) and (\ref{eq:M12sHiggs}). 
In these cases the RG effects are substantially reduced and  
the chiral enhancement is basically absent. One finds 
\be
P_2^{LR} (B_d) \approx P_2^{LR} (B_s) \approx 3.4 \qquad {\rm for} \qquad  \mu_H=246~{\rm GeV}.
\ee 
This is about a factor of four larger
than the corresponding factor for the SM operator: $P_1^{VLL} (B) \approx 0.72$.
We then conclude that the bounds on $(\widetilde \Delta_{d})_{31}$ and
$(\widetilde \Delta_{d})_{32}$ are substantially weaker than the bound 
coming from $\epsK$ of Eq.~(\ref{eq:epsdbound}). 
 
For completeness, we report here the bounds 
on the effective FCNC scalar couplings 
assuming a NP contribution, in magnitude,
up to $20\%$ ($50\%$) of the SM
$B_d$ ($B_s$) mixing amplitude: 
\bea
| \epsilon_d | \times \left|(\widetilde \Delta_{d})^*_{31} (\widetilde \Delta_{d})_{13})\right|^{1/2}
  &\lsim&  5 \times 10^{-5}  \times  \frac{ c_\beta M_H}{100~{\rm GeV}}~,
\label{eq:Delta13} \\
| \epsilon_d | \times \left| (\widetilde \Delta_{d})^*_{32} (\widetilde \Delta_{d})_{23}\right|^{1/2}
  &\lsim&  3 \times 10^{-4}  \times  \frac{ c_\beta M_H}{100~{\rm GeV}}~.
\label{eq:Delta23}
\eea  
Note that if we enforce the RS-GIM structure in 
(\ref{eq:RSGIM}) and the bound on $\epsilon_d$ in 
(\ref{eq:RSGIMbound}), derived from $\epsK$,
we obtain theoretical constraints on the effective FCNC
couplings which are well below the phenomenological 
bounds in (\ref{eq:Delta13}) and  (\ref{eq:Delta23}):
\bea
| \epsilon_d | \times \left|(\widetilde \Delta_{d})^*_{31} (\widetilde \Delta_{d})_{13})\right|^{1/2}_{\rm RS-GIM}
  &\lsim&  2 \times 10^{-6}  \times  \frac{ c_\beta M_H}{100~{\rm GeV}}~, \\
| \epsilon_d | \times \left| (\widetilde \Delta_{d})^*_{32} (\widetilde \Delta_{d})_{23}\right|^{1/2}_{\rm RS-GIM}
  &\lsim&  1 \times 10^{-5}  \times  \frac{ c_\beta M_H}{100~{\rm GeV}}~.
\eea
In other words, as anticipated, within the RS-GIM framework
the $\epsK$ constraint naturally forbids significant effects 
in the $B_{s,d}$ systems.

\subsection{\boldmath  Neutral-Higgs exchange in $\Delta F=2$ amplitudes: 
$\text{2HDM}_{\overline{\text{MFV}}}$}
\subsubsection{Basic formulae} 
Using the effective Hamiltonians in Section~\ref{sect:fbphases} and proceeding like in the general case we find
\bea
\label{eq:M12KHiggsMFV}
M_{12}^K &=& -\frac{1}{3}M_K F_K^2 P_2^{LR}(K)
\frac{|a_0|^2}{  M_H^2} ~y_s y_d [y_t^2 V_{ts} V^*_{td} ]^2~, \\
\label{eq:M12dHiggsMFV}
M_{12}^d &=& -\frac{1}{3}M_{B_d} F_{B_d}^2 P_2^{LR}(B_d)
\frac{(a_0+a_1)(a^*_0+a^*_2)}{  M_H^2} ~y_b y_d [y_t^2 V_{tb} V^*_{td} ]^2, \\
\label{eq:M12sHiggsMFV}
M_{12}^s &=& -\frac{1}{3}M_{B_s} F_{B_s}^2 P_2^{LR}(B_s)
\frac{(a_0+a_1)(a^*_0+a^*_2)}{  M_H^2} ~y_b y_s [y_t^2 V_{tb} V^*_{ts} ]^2~.
\eea
For comparison, we recall that the SM contribution to $M_{12}^q$ reads 
\be
\left(M_{12}^q\right)_\text{SM} = \frac{G_F^2}{12 \pi^2} M_W^2 (V_{tb} V_{tq}^*)^2  F^2_{B_q} 
M_{B_q} \eta_B\hat B_{B_q}(B_q) S_0^*(x_t)~,
\ee
with the loop function $S_0^*(x_t=m_t^2/M_W^2)$\footnote{In spite of the fact that the function $S_0$ is real, we put a $*$ on this function, anticipating a new phase in the effective $S$ function we will obtain in presence of NP effects (see next section).} 
given in \cite{Buras:2001ra}. 
A similar expression holds for the SM top-quark contribution
to $M_{12}^K$, changing accordingly the flavour indices and with the replacement $\eta_B\hat B_{B_q}\to\eta_{tt} \hat B_{K}$ (see Ref.~\cite{Buras:2001ra} for notations).

\subsubsection{\boldmath  Phenomenology of  the $K^0$--$\bar{K}^0$ and  $B^0_{s,d}$--$\bar{B}^0_{s,d}$ systems}

Using the basic formulae of above it is
straightforward to find the expressions for $\epsK$, the mass differences
$\Delta M_q$ and the asymmetries $S_{\psi K_S}$ and $S_{\psi\phi}$.

In the case of $\epsK$ the inclusion of neutral Higgs 
contributions amounts to the replacement of the SM box function $S_0(x_t)$ by
\be\label{eq:SK}
S_K=S_0(x_t)-\frac{64\pi^2}{\hat B_K \eta_{tt} } P_2^{LR}(K)
|a_0|^2~\frac{m_d m_s}{M_W^2} \frac{m_t^4 t^2_\beta}{M_H^2v^2}~,
\ee
where we approximated $1/(c_\beta^2s_\beta^4)$ by $t^2_\beta$.

In order to write down analogous expressions for the box functions $S_q$
relevant for the $B^0_{s,d}$--$\bar{B}^0_{s,d}$ systems, we introduce
\be
T_q=\frac{64\pi^2}{\hat B_{B_q} \eta_B} P_2^{LR}(B_q)
(a^*_0+a^*_1)(a_0+a_2)\frac{m_b m_q}{M_W^2} \frac{m_t^4 t_\beta^2}{M_H^2v^2}~,
\ee
where $T_s$ and $T_d$ are unambiguously connected by the relation
\be
T_d ~=~ \frac{m_d {\hat B}_{B_s}  P_2^{LR}(B_d) }{m_s {\hat B}_{B_d} P_2^{LR}(B_s) }~ T_s
~\approx~ \frac{m_d}{m_s} ~T_s~.
\label{eq:TdTs}
\ee
Then we simply  have 
\be
\label{eq:Sq}
 S_q=S_0(x_t)-T_q =|S_q|e^{-2i\varphi_{B_q}}~,
\ee
$S_{\psi K_S}$ and $S_{\psi\phi}$ given in (\ref{eq:3.43}) and
\label{eq:NewDMq}
\bea
&& \Delta M_q=\frac{G_F^2}{6\pi^2}\eta_B M_{B_q}\hat B_{B_q}F^2_{B_q}M_W^2|V_{tq}|^2
|S_q|~.
\eea
From (\ref{eq:TdTs}) it follows that
\be
\varphi_{B_d}\approx \frac{m_d}{m_s}\varphi_{B_s}.
\ee

\begin{figure}[t]
\begin{center}
\includegraphics[width=.45\textwidth]{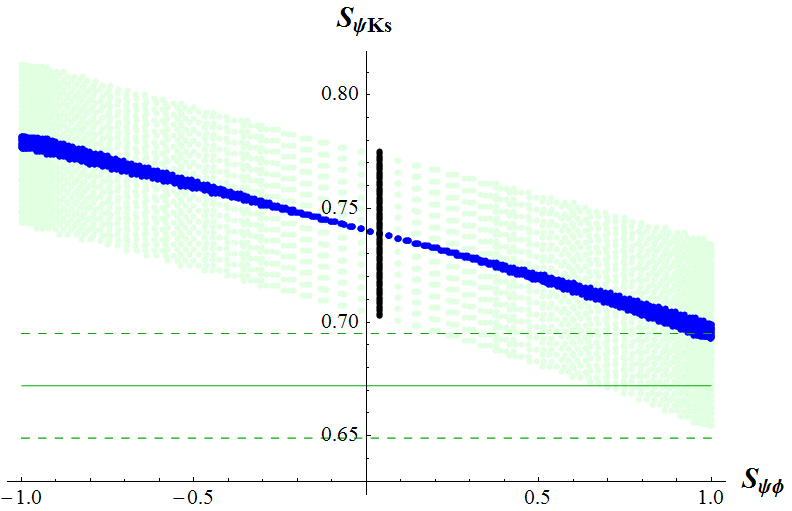}
\includegraphics[width=.45\textwidth]{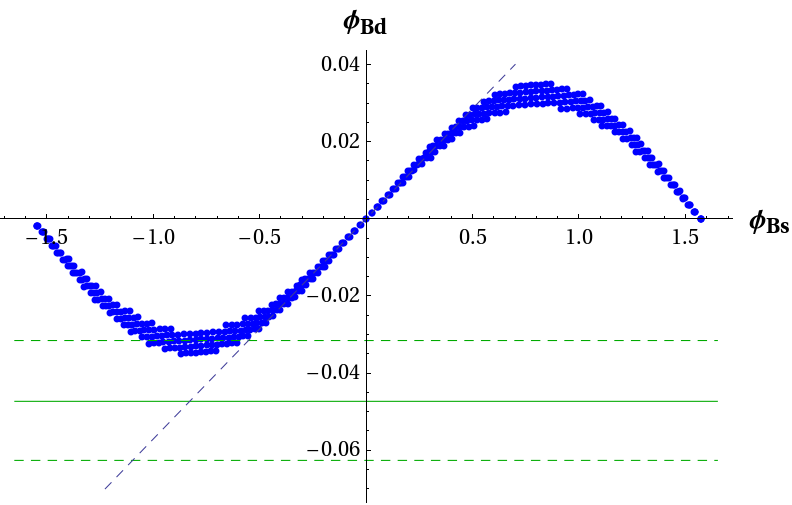}
\vskip 0.5 true cm 
\caption{\label{fig:sin2b}
Correlation between $S_{\psi K_S}$ and $S_{\psi\phi}$ (left) and between the new phases in the
$B_d$ and $B_s$ mixing (right) originating in the
CP-violating Higgs-mediated $\Delta F=2$ amplitudes in Eqs.~(\ref{eq:M12dHiggsMFV})--(\ref{eq:M12sHiggsMFV}). 
In both plots the blue (dark) points
have been obtained with the CKM phase $\beta$ fixed to its central value ($\sin(2\beta)=0.739$): 
the spread is determined only by the condition imposed on $\Delta M_s$
(see text). The horizontal lines indicate the $\pm1\sigma$ range of $S_{\psi K_S}^{\text{exp}}$.
On the left plot the $\pm 1\sigma$ error due to the uncertainty in the extraction of 
$\beta$ (light points) and the SM prediction (black vertical line) are also shown. The dashed blue (dark) line in the right plot represents $\phi_{B_s} = \left( m_d/m_s \right) \phi_{B_s}$. }
\end{center}
\end{figure}

\begin{figure}[t]
\begin{center}
\includegraphics[width=.6\textwidth]{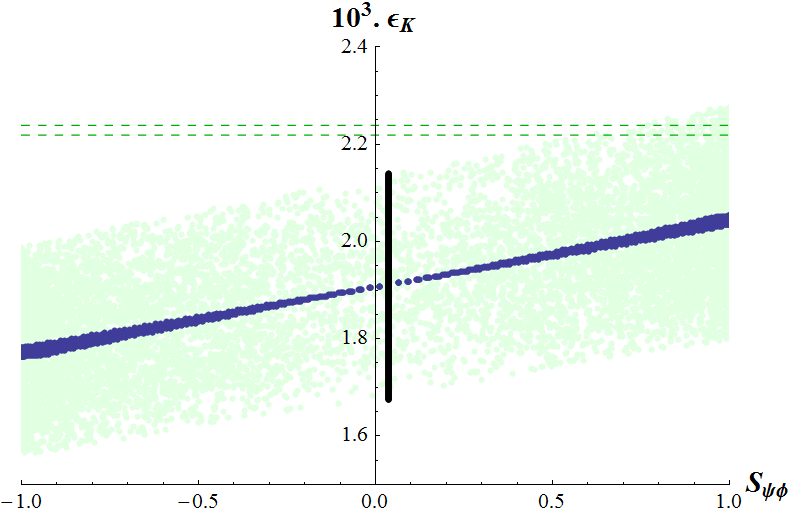}
\vskip 0.5 true cm 
\caption{\label{fig:epsK}
Correlation between $\epsK$ and $S_{\psi K_S}$  with the inclusion of the 
CP-violating Higgs-mediated $\Delta F=2$ amplitudes in Eqs.~(\ref{eq:M12dHiggsMFV})--(\ref{eq:M12sHiggsMFV}).
Notations as in Figure~\ref{fig:sin2b}.}
\end{center}
\end{figure}

Before presenting the numerical analysis of these formulae, we would like to
list the most relevant properties of these results, some of which have 
already been outlined in Section~\ref{sect:fbphases}:
\begin{itemize}
\item
 The flavour universality of the box function $S$ is broken, with
 violations governed by the quark masses relevant for the particular
 system considered: $m_sm_d$, $m_b m_d$ and $m_b m_s$, 
for the $K$, $B_d$ and $B_s$ systems, respectively. This opens the 
 possibility of a large impact in $B_s$ mixing solving the problem of the 
 large CP-violation phase hinted by  by CDF~\cite{Aaltonen:2007he} 
 and D0~\cite{Abazov:2008fj,Abazov:2010hv}. However, this can happen only 
 with large flavour blind phases, otherwise, with $a_i$ real, the CP 
 asymmetries would not be affected. 
 \item
 If we try to accommodate a large CP-violating phase in $B_s^0$--$\bar B_s^0$ mixing
 in this scenario, we find a correlated shift in the relation between $S_{\psi K_S}$ and the CKM phase $\beta$. 
 This shift is determined unambiguously by the relation between $T_d$ 
 and $T_s$ in Eq.~(\ref{eq:TdTs}): it contains no free parameters. The shift is such that the
 prediction of $S_{\psi K_S}$ {\em decreases} with respect to the SM case 
 at fixed CKM inputs (assuming a large positive value of $S_{\psi \phi}$, 
 as hinted by CDF and D0). This relaxes the existing 
 tension between  $S^{\rm exp}_{\psi K_S}$ and its SM prediction (see Figure~\ref{fig:sin2b}).
 \item
 The new physics contribution to $\epsK$ is tiny and has unique sign, implying a 
 destructive interference with the SM box amplitude. This contribution 
 alone does not improve the agreement between data and prediction for $\epsK$. 
 However, given the modified relation between 
 $S_{\psi K_S}$ and the CKM phase $\beta$, the true value of $\beta$ extracted in this 
 scenario increases with respect to SM fits. As a result of this modified value of 
 $\beta$, also the predicted value for $\epsK$ {\it increases} with respect to 
 the SM case, resulting in a better agreement with data (see Figure~\ref{fig:epsK}).
 \end{itemize}

\noindent
The numerical analysis of these effects, illustrated in Figure~\ref{fig:sin2b} and~\ref{fig:epsK},
has been obtained adopting the following two strategies: 
\begin{description}
\item[{\bf Figure~\ref{fig:sin2b}:}] 
We combine the value of $\sin(2\beta)$ determined by tree-level
observables (see Table~\ref{tab:eps}) with $\epsK$ (assuming negligible new-physics 
effects in $\epsK$) finding a reference value of $\sin(2\beta)$ independent from $S^{\rm exp}_{\psi K_S}$.
The result thus obtained is $\sin(2\beta) = 0.739 \pm 0.036$. 
We determine the size of $T_s$, as a function of $\varphi_{B_s}$,
requiring a deviation of $\Delta M_s$ within $10\%$ of its SM value. This fixes unambiguously all
our free parameters as function of the CP-violating phase of the $B_s$ mixing amplitude 
(or the CP-asymmetry $S_{\psi \phi}$). 
The $B_d$ mixing phase and the  CP-asymmetry $S_{\psi K_S}$ are then computed by means of 
(\ref{eq:3.43}) in terms of $\varphi_{B_s}$ and the reference value of $\beta$.
\item[{\bf Figure~\ref{fig:epsK}:}] 
For each value of the $B_s$ mixing amplitude we determine the value of the new-physics phase 
 $\varphi_{B_s}$. Also in this case the magnitude of $T_s$ is fixed requiring 
a deviation of $\Delta M_s$ within $10\%$ of its SM value.
By means of (\ref{eq:3.43}) we then extract  $\sin(2\beta)$ from  $S^{\rm exp}_{\psi K_S}$
obtaining a reference value of $\sin(2\beta)$ independent from $\epsK$. 
The value of  $\epsK$ is then predicted using SM expressions  as a function 
of the new reference value of $\sin(2\beta)$ (which in turn depends 
on the $B_s$ mixing phase).
\end{description}

By looking at the plots in  Figure~\ref{fig:sin2b} and~\ref{fig:epsK} it is quite clear 
that a large positive value of $S_{\psi \phi}$ (or a large negative $\varphi_{B_s}$), as hinted by 
by CDF~\cite{Aaltonen:2007he} and D0~\cite{Abazov:2008fj,Abazov:2010hv}, can easily be explained
in this framework and, more important, this implies, as a byproduct, a substantial improvement 
in the predictions of $S_{\psi K_S}$ and $\epsK$.

In order to understand for which range of the underlying model parameters the desired 
effect is produced, we list here the conditions of negligible direct impact 
on $\epsK$ and sizable contribution to $B_s$ mixing. 
The direct impact of the Higgs-mediated amplitude in $\epsK$ do not exceed the $5\%$
level for
\be
|a_0| t_\beta \frac{v}{M_H } < 18~,
\ee
while for 
\be
\sqrt{|(a^*_0+a^*_1)(a_0+a_2)|} t_\beta \frac{v}{M_H } = 10 \qquad{\rm and}\qquad  
\arg{[(a^*_0+a^*_1)(a_0+a_2)]} \approx -1.2
\label{eq:aicond}
\ee
we get $S_{\psi \phi} \approx 0.4$, with $\Delta M_s$
within $10\%$ of its SM value.
As can be noted, the two conditions are perfectly compatible.\footnote{~A large 
contribution to $B_s^0$--$\bar B_s^0$ mixing is also compatible with the bounds on $t_\beta$ 
and $M_H$ derived by the charged-Higgs exchange in 
$B\to\tau\nu$ (see e.g.~\cite{Bona:2009cj}). These bounds 
are almost independent from the $a_i$ and can easily be satisfied 
for $t_\beta = \cO(10 \times M_H/v)$. Incidentally, we note that the prediction 
of $B\to\tau\nu$ could even improve in this framework 
because of the higher value of $|V_{ub}|$ extracted from 
the global analysis of the CKM unitarity triangle, when the Higgs-mediated $\Delta B=2$
amplitude is taken into account.}
The range of free parameters is also very natural, with $a_i$ of order one
and $t_\beta$ moderate ($t_\beta\sim 10$) or large ($t_\beta\sim 50$)
depending on the value of $M_H$. The only condition which is not trivial 
to achieve is the large CP-violating phase. The latter requires a large difference 
between $a_1$ and $a_2$ that, as already pointed out in 
Section~\ref{sect:fbphases}, is not easy to obtain in 
explicit new physics models.

\subsection{\boldmath  The rare decays  $B_{s,d} \to \mu^+\mu^-$}

The rare decays $B_{s,d} \to \mu^+\mu^-$ could provide the 
ultimate and decisive test about magnitude and flavour structure
of Higgs-mediated FCNC amplitudes. 

Using the effective Hamiltonian in (\ref{eq:HeffBll}) and taking into account 
the known SM contribution we find
\be
{\rm Br}(B_q\to\mu^+\mu^-)= {\rm Br}(B_q\to\mu^+\mu^-)_{\rm SM}\times\left( |1+R_q|^2 + |R_q|^2 \right)~,
\ee
where 
\be
R_q = (a_0^*+a^*_1) ~\frac{2\pi^2 m_t^2}{Y_0(x_t) M_W^2} ~\frac{M_{B_q}^2 t_\beta^2}{(1+m_q/m_b) M^2_H}~, \label{eq:85}
\ee
and 
\be
{\rm Br}(B_q\to\mu^+\mu^-)_{\rm SM} 
   =  \frac{G_F^2\tau_{B_q}}{\pi} \left(\frac{g^2}{16\pi^2}\right)^2
  F^2_{B_q } M_{B_q } m^2_\ell 
    \sqrt{1-\frac{4 m^2_\ell}{M^2_{B_q }}} \left|V_{tb}^*V_{tq}\right|^2 Y(x_t)^2~,
\ee
with the loop function $Y(x_t)$ given, for instance, in~\cite{Altmannshofer:2009ne}. 
In the case of the SM the relation of ${\rm Br}(B_q\to\mu^+\mu^-)$ to
$\Delta M_q$ pointed out in~\cite{Buras:2003td} allows to reduce the uncertainty and one 
finds
\bea
{\rm Br}(B_d\to\mu^+\mu^-)_{\rm SM} &=& (1.0\pm 0.1)\times 10^{-10}\,,  \nn \\
{\rm Br}(B_s\to\mu^+\mu^-)_{\rm SM} &=& (3.2\pm 0.2)\times 10^{-9}\,.
\eea

\begin{figure}[t]
\begin{center}
\includegraphics[width=.8\textwidth]{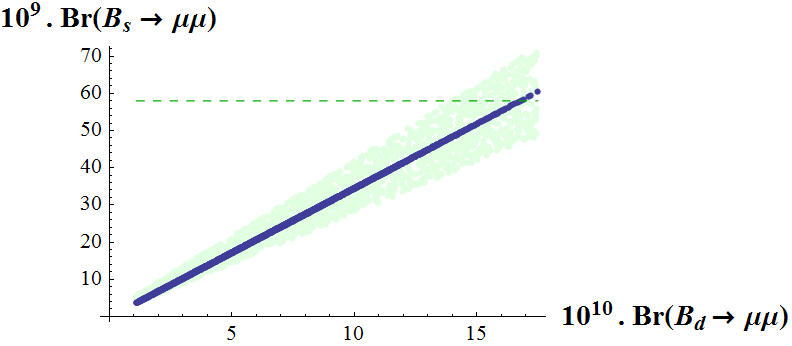}
\vskip 0.5 true cm 
\caption{\label{fig:Bll}
Correlation between ${\rm Br}(B_s\to\mu^+\mu^-)$ and ${\rm Br}(B_d\to\mu^+\mu^-)$
in presence of scalar amplitudes respecting the MFV hypothesis.
The horizontal dotted line represent the present experimental
limit on ${\rm Br}(B_s\to\mu^+\mu^-)$ from Ref.~\cite{:2007kv}.}
\end{center}
\end{figure}

The striking feature of (\ref{eq:85})
 is the almost exact universality of $R_s$ and $R_d$:
their difference, due to light-quark masses, is well below the parametric 
uncertainties on the SM predictions for the two branching ratios.
This universality, which is not affected by the presence of possible flavour-blind phases,
leads to the strict correlation shown in Figure~\ref{fig:Bll}. 
This correlation
holds not only for Higgs-mediated amplitudes but, more generally, 
in the presence of both scalar and $V-A$ operators with a MFV 
structure~\cite{Hurth:2008jc}. It can indeed be 
considered as a ``smoking-gun'' of the MFV hypothesis~\cite{Hurth:2008jc,Blanke:2006ig}.

Eliminating the dependence of ${\rm Br}(B_{s} \to\mu^+\mu^-)/{\rm Br}(B_{d} \to\mu^+\mu^-)$
from $|V_{ts}/V_{td}|$ in terms of 
the corresponding dependence of $\Delta M_{s}/\Delta M_{d}$  we can write
\be\label{eq:r}
\frac{{\rm Br}(B_s\to\mu^+\mu^-)}{{\rm Br}(B_d\to\mu^+\mu^-)}= \frac{\hat
B_{B_d}}{\hat B_{B_s}} \frac{\tau(B_s)}{\tau(B_d)} \frac{\Delta
M_s}{\Delta M_d}\,r\,,\quad r= \frac{M_{B_s}^4}{M_{B_d}^4}
\frac{|S_d|}{|S_s|}\,,
\ee
where for $r=1$ we recover the SM and CMFV relation 
derived in~\cite{Buras:2003td}. 
In our general MFV framework 
$r$ can deviate from one; however, this deviation is at most of $\cO(10\%)$,
as outlined in the previous section. Actually a precise 
measurement of the two $B_{s,d} \to\mu^+\mu^-$
rates would be the best way to determine $r$ and, by means 
of (\ref{eq:r}), the amount of non-standard contributions to 
$\Delta M_{s}/\Delta M_{d}$.

This strict correlation of  ${\rm Br}(B_{s,d} \to\mu^+\mu^-)$  shown in  Figure~\ref{fig:Bll}
should be contrasted with non-MFV frameworks, such as the MSSM with non-minimal 
flavour structures~\cite{Isidori:2002qe,Chankowski:2000ng,Altmannshofer:2009ne}
or models with warped space-time geometry~\cite{Blanke:2008yr}.
In some of these frameworks large enhancements of the two ${\rm Br}(B_{s,d}\to\mu^+\mu^-)$
are possible, but their ratio is no more related to 
 $|V_{ts}/V_{td}|^2$. As a result, the plots corresponding 
to Figure~\ref{fig:Bll} look very differently
(see in particular the plots in~\cite{Altmannshofer:2009ne}).
  
The upper limit ${\rm Br}(B_s\to\mu^+\mu^-) < 5.8\times 10^{-8}$~\cite{:2007kv} implies 
\be
\sqrt{|a_0+a_1|} t_\beta \frac{v}{M_H } < 8.5~.
\ee
This result is compatible with (\ref{eq:aicond}) only if the $a_i$ are 
of order 1 (for $a_i \ll 1$ it would require an unnaturally large value for $a_2/a_1$).
For $a_i=\cO(1)$ it signals that ${\rm Br}(B_s\to\mu^+\mu^-)$ has to be close 
to its experimental  bound if scalar amplitudes play a significant role in $B_s$ mixing.

\section{Conclusions}

In the present paper we have addressed the question of the effectiveness 
of the NFC and MFV hypotheses in suppressing the strength of FCNC transitions 
to the observed level in models with more than one-Higgs doublet.
More generally, we have analysed the interplay between continuous and discrete
flavour-blind symmetries and the symmetry-breaking pattern in the 
flavour sector, in determining the structure of scalar FCNCs. We have 
considered explicitly only a general 2HDM, but our discussion applies also 
to models with more than two-Higgs doublets. 

The NFC hypothesis is based on the imposition of flavour-blind symmetries 
in the Yukawa Lagrangian which do not hold beyond the tree-level. 
On the contrary, the 
MFV hypotheses is based on a symmetry and symmetry-breaking pattern in the 
flavour sector which is renormalization group invariant. As a result, it is 
not surprising that the latter framework turns out to be superior. 
As we have shown, this is evident when effects beyond 
the tree-level are taken into account.
As we have illustrated with the example of $\epsK$, beyond the tree-level 
the NFC hypothesis ceases to provide a sufficient suppression of FCNCs in a natural 
manner, unless the neutral Higgs masses are well above the LHC energy scales. 
This should be contrasted with MFV framework, which is stable under higher 
order contributions. Within the latter framework relatively low values 
for the Higgs masses can be accommodated, as well as large values of the bottom
Yukawa coupling (namely $t_\beta \gg 1$). 
The comparison between NFC and MFV has also given us the opportunity 
to clarify which of the multi-Higgs models proposed in the recent literature
are consistent with the MFV principle, and thus are naturally 
protected against too large FCNCs.

The MFV hypothesis is very simple, theoretically sound and, as we have just 
stressed, very efficient in suppressing large FCNC contributions to the 
measured level. Yet, the important phenomenological question remains 
whether such a constrained framework is at the end consistent with the 
nature around us. Indeed, there exist at least three anomalies observed in
the data that, at first sight, give a clear hint for the presence of non-MFV 
interactions. These are as follows.
\begin{itemize}
\item
First of all the size of the CP-violation in $B_s^0$--$\bar B_s^0$ system 
signaled by the CP-asymmetry $S_{\psi\phi}$ in $B_s\to \psi\phi$ observed by
CDF and D0 that appears to be  roughly by a factor of 20 larger than the 
SM and MFV predictions, assuming the Yukawa couplings to be the only sources 
of CP-violation.
\item
The value of $\sin 2\beta$ resulting from the UT fits tends to be significantly 
larger than the measured value of $S_{\psi K_S}$.
\item
The value of $\epsK$ predicted in the SM by using $S_{\psi K_S}$ as the measure of 
the observed CP-violation is about $2\sigma$ lower than 
the data. In short, the values of $S_{\psi K_S}$ and $\epsK$ cannot be 
simultaneously described within the SM~\cite{Lunghi:2008aa,Buras:2008nn}.
\end{itemize}
As pointed out in Ref.~\cite{Kagan:2009bn,Mercolli:2009ns,Paradisi:2009ey}, 
the mechanisms of flavour and CP-violation do not necessarily need to 
be related. In particular, as noted in \cite{Kagan:2009bn}, 
a large new phase in $B^0_s$--$\bar B^0_s$ mixing could
in principle be obtained in the MFV framework if additional flavour-blind
phases are present. This idea cannot be realized in the ordinary MSSM
with MFV, as shown in~\cite{Altmannshofer:2009ne}.\footnote{~As discussed in 
Section~\ref{sect:fbphases}, the difficulty of realizing this scenario in the MSSM 
is due to the suppression in the MSSM 
of effective operators with several Yukawa insertions. 
Sizable couplings for these operators are necessary both to have an 
effective large CP-violating phase in $B^0_s$--$\bar B^0_s$ mixing and, at the same time,
 to evade bounds from 
other observables, such as $B_s\to \mu^+\mu^-$ {and $B \to X_s \gamma$.}}
{However, it could be realized in different underlying models, 
such as the up-lifted MSSM, as recently pointed out in Ref.~\cite{Dobrescu:2010rh}.}  

In the present work we have demonstrated that a general 2HDM with MFV, enriched by 
flavour blind phases ($\text{2HDM}_{\overline{\text{MFV}}}$), is not only capable in explaining the first anomaly (large $S_{\psi\phi}$): 
once the first problem is addressed,  unique solutions to the other
two problems  listed above ($S_{\psi K_s}$--$\varepsilon_K$) are naturally at work.
Indeed, a small new phase in $B_d^0$--$\bar B_d^0$ system 
with the correct sign and roughly correct size is automatically implied 
by a large phase in the $B_s^0$--$\bar B_s^0$ system with the hierarchy of these
two new phases being fixed by the ratio $m_d/m_s$. In practice this solves 
(or at least softens) the remaining
two problems by enhancing the true value of the phase $\beta$ of $V_{td}$:
the values of $S_{\psi K_s}$ and $\epsK$ thus obtained 
are in better agreement with expectations in spite of the negligible contribution 
of neutral Higgs exchanges to $\epsK$. In summary, the neutral Higgs exchanges 
contributing to $\Delta F=2$ processes in the $\text{2HDM}_{\overline{\text{MFV}}}$
seems to be an interesting solution 
to all these anomalies, with a clear pattern of correlations that could be 
easily verified or falsified in the near future.

It must be stressed that this mechanisms and pattern of correlations are quite 
different than other mechanisms proposed in the recent literature to accommodate a
large $B_s^0$--$\bar B_s^0$ mixing phase.
Those models typically contain more free parameters (associated to 
new flavour-breaking sources) and do not provide a natural explanation of 
why the deviations from the SM should be small (or vanishingly small) 
in  the $B_d^0$--$\bar B_d^0$ mixing and in $\epsK$.

Finally, we have stressed the key role of the rare decays $B_{s,d} \to \mu^+\mu^-$ 
in providing a decisive test of the flavour-breaking structure
implied by MFV, independently of possible flavour-blind phases, 
and independently of the dominance of scalar vs.~vector
FCNC operators beyond the SM.

\section*{Acknowledgments}
We thank Wolfgang Altmannshofer, Monika Blanke,
and Toni Pich for useful comments and discussions.
AJB would like to thank Gustavo Branco for bringing to his attention 
Ref.~\cite{Botella:2009pq}, which motivated us the 
reconsideration of MFV in multi-Higgs models. 
AJB, SG, and GI  thank the Galileo Galilei Institute for Theoretical Physics 
for the hospitality and partial support during the completion of this work. 
This research was partially supported by the Cluster of Excellence `Origin and Structure 
of the Universe', by the Graduiertenkolleg GRK 1054 of DFG,
by the German `Bundesministerium f\"ur Bildung und Forschung' under 
contract 05H09WOE, and by the EU Marie Curie Research Training Network
contracts MTRN-CT-2006-035482 ({\em Flavianet})
and MRTN-CT-2006-035505 ({\em HEP-TOOLS}).

\section*{Added note}
After the completion of this work, two papers discussing related issues
have appeared: 
a model-independent analysis of new-physics effects in $B_s$ 
and $B_d$ mixing~\cite{Ligeti:2010ia}, 
and a detailed analysis of flavour-changing 
amplitudes mediated by charged-Higgs exchange~\cite{Jung:2010ik}.

\section*{Appendix}
The Higgs Lagrangian of a generic  model with two-Higgs doublets, $H_1$ and 
$H_2$, with hypercharges $Y=1/2$ and $Y=-1/2$ respectively, can be decomposed as
\be
\cL^{\rm 2HDM}_{\rm Higgs} = \sum_{i=1,2} D_\mu H_i D_\mu H_i^\dagger + 
\cL_{Y}  - V(H_1,H_2)~,
\ee
where $D_\mu H_i = \partial_{\mu}H_i-i g^\prime Y \hat{B}_{\mu}H_i -ig T_a\hat{W^a}_{\mu} H_i$, with $T_a=\tau_a/2$. 
The potential is such that the $H_i$ gets a non-trivial vev, giving
rise to non-vanishing masses for $M_W$ and $M_Z$ bosons. In the unitary gauge
we can set 
\be
\langle H_{1} \rangle = \frac{1}{\sqrt{2}} \left(\ba{c} 0 \\ v_1 \ea\right)~,\hspace{0.2cm} \langle H_{2} \rangle = \frac{1}{\sqrt{2}} \left(\ba{c} v_2 \\  0  \ea\right)~,
\qquad  v^2 =v_1^2+v_2^2 \approx 246~\left( \text{GeV} \right)^2~,
\ee
where $v_1$ and $v_2$ can be always taken positive, and we assume their phases to be zero in order to avoid spontaneous CP breaking.

A useful change of basis is obtained with the following global rotation 
	\be
	\begin{pmatrix} \Phi_v \\ \Phi_H \end{pmatrix} = \frac{1}{v} \begin{pmatrix} v_1 & v_2 \\ -v_2 & v_1 \end{pmatrix} \begin{pmatrix} H_1 \\ H^c_2 \end{pmatrix}~.
\label{eq:tbrotation}
	\ee
In the new basis only the doublet $\Phi_v$ has a non-vanishing vev, and the eight degrees of freedom of the two-Higgs doublets appear explicitly as three Goldstone bosons $G^{\pm}$ and $G^0$, two charged Higgs $G^{\pm}$, and three neutral scalars $S_{1,2,3}$:
	\be\label{eq:higgsbasis}
	\Phi_v =  \begin{pmatrix} G^+ \\\frac{1}{\sqrt{2}}( v + S_1 + i G^0) \end{pmatrix} ~,
	\qquad \qquad
	\Phi_H = \begin{pmatrix}  H^+\\\frac{1}{\sqrt{2}} (S_2 + i S_3)  \end{pmatrix}~.
	\ee 

It is worth to stress that, in the absence of an interaction distinguishing the 
two-Higgs fields, the parameter $t_\beta\equiv\tan(\beta) = v_2/v_1$ is not
well defined: we can always mix the two fields by means of rotations of the type
(\ref{eq:tbrotation}). This is not the case if we assume 
an exact $U(1)_{\rm PQ}$ symmetry (or its discrete $Z_2$ subgroup),
which would allows us to distinguish the two fields.
Throughout this paper we assume the $U(1)_{\rm PQ}$ breaking terms in the 
Higgs potential are calculable (or identifiable) from first principles 
(e.g.~the $U(1)_{\rm PQ}$ symmetry is only softly broken, as in the MSSM), 
such that the two fields can be defined starting from the 
$U(1)_{\rm PQ}$ limit of the theory (a detailed discussion about the definition 
of $t_\beta$ beyond the tree-level in the MSSM can 
be found in Ref.~\cite{Gorbahn:2009pp}).

The most general potential for the two-Higgs doublets that is renormalizable
and gauge invariant is
	\begin{eqnarray}\nonumber 
	V(H_1,H_2)&=&\mu_1^2|H_1|^2+\mu_2^2|H_2|^2+(b H_1 H_2+{\rm h.c})+\frac{\lambda_1}{2}|H_1|^4+\frac{\lambda_2}{2}|H_2|^4+\lambda_3|H_1|^2|H_2|^2\\
	&+&\lambda_4|H_1H_2|^2+\left[\frac{\lambda_5}{2}(H_1H_2)^2+\lambda_6|H_1|^2H_1H_2+\lambda_7|H_2|^2H_1H_2+{\rm h.c}\right]\,,
	\end{eqnarray}
where $H_1 H_2= H_1^T (i \sigma_2) H_2$ and 
all the parameters must be real with the exception 
of $b$ and $\lambda_{5,6,7}$. Changing the relative phase of $H_1$ and $H_2$ 
we can cancel the phase of $b$ and $\lambda_{6,7}$ relative to $\lambda_{5}$. 
As a result, the potential is invariant under CP if $b$ and $\lambda_{5,6,7}$
are all real, if only one of these couplings is different from zero, 
or if their phases are related [arg($\lambda_5$)=2~arg($b$)=2~arg($\lambda_{6,7}$)].

In several explicit models the coefficients $\lambda_{6,7}$ are set to zero.
This can be achieved by imposing a discrete $Z_2$ symmetry
that is only softly broken by the terms proportional to $b$ and $\lambda_5$. 
Both $b$ and $\lambda_5$ break the  $U(1)_{PQ}$ symmetry
and, if $\lambda_6=\lambda_7=0$,  at least one of them  must be non-zero 
to prevent the appearance of a massless pseudoscalar Goldstone boson.

\medskip

In order to analyse the spectrum of the theory, let us first restrict the attention to 
the case of exact CP invariance. In this case the neutral mass eigenstates are 
two CP-even ($h^0$ and $H^0$) and one CP-odd ($A^0$) states. 
The masses of the CP-odd and charged fields are
	\begin{align}
	M_A^2 &= \frac{b}{s_{\beta} c_{\beta}} -\frac{1}{2} v^2 \left( 2 \lambda_5 - \lambda_6 t_{\beta}^{-1} - \lambda_7 t_{\beta} \right)~, \\
	M_{H^{\pm}}^2 &= M_{A}^2 + \frac{1}{2} v^2 \left( \lambda_5 - \lambda_4 \right)~.
	\label{massA}
	\end{align}
The two CP-even states mix according with the squared-mass matrix
	\be
	\mathcal{M}^2 = M_A^2 \begin{pmatrix} s_{\beta}^2 & - s_{\beta} c_{\beta} \\ - s_{\beta} c_{\beta} & c_{\beta}^2 \end{pmatrix} + \mathcal{B}^2~,
	\ee
        \be
	\mathcal{B}^2 = \begin{pmatrix} \lambda_1 c_{\beta}^2 + \lambda_5 s_{\beta}^2 - 2 \lambda_6 s_{\beta} c_{\beta} & \left( \lambda_3 + \lambda_4 \right) s_{\beta} c_{\beta} - \lambda_6 c_{\beta}^2 - \lambda_7 s_{\beta}^2 \\  
\left( \lambda_3 + \lambda_4 \right) s_{\beta} c_{\beta} - \lambda_6 c_{\beta}^2 - \lambda_7 s_{\beta}^2 & \lambda_2 s_{\beta}^2 + \lambda_5 c_{\beta}^2 - 2 \lambda_7 s_{\beta} c_{\beta} \end{pmatrix}~.
	\ee
whose eigenstates are
	\bea
&&	\begin{pmatrix} H^0 \\ h^0 \end{pmatrix} =  \begin{pmatrix} \cos(\alpha-\beta) & \sin(\alpha-\beta) \\ -\sin(\alpha-\beta) & \cos(\alpha-\beta) \end{pmatrix} \begin{pmatrix} S_1 \\ S_2 \end{pmatrix}~, \\
&&       m^2_{H^0,h^0} = \frac{1}{2} \left[ \mathcal{M}^2_{11} + \mathcal{M}^2_{22} \pm \sqrt{ \left( \mathcal{M}^2_{11} - \mathcal{M}^2_{22} \right)^2 + 4 \left( \mathcal{M}^2_{12} \right)^2} \right]~,
	\eea
The CP-even eigenstates
are defined such that $M_{h^0} < M_{H^0}$, and the explicit expression of the 
mixing angle $\alpha$ is
	\be
	\tan({2 \alpha}) = \frac{2 \mathcal{M}^2_{12}}{\mathcal{M}^2_{11} - \mathcal{M}^2_{22}}~. \qquad 
		\ee
As discussed in~\cite{GunionHaber}, for $ M_A^2 \gg | \lambda_i | v^2 $ we are 
in the decoupling regime where 
\be	
M_{h^0} = \mathcal{O} (v)~, \qquad 
M_{H^0},  M_{H^{\pm}} = M_A  +  \mathcal{O} \left(\frac{v^2}{M_A}\right)~, \qquad 	
\cos({\beta-\alpha}) = \mathcal{O} \left(\frac{v^2}{M_A^2}\right)~.
\ee
This regime is characterised by a negligible
mass mixing of the two doublets $\Phi_v$ and $\Phi_H$, 
and  by two separate mass scales ($M_A \gg M_{h^0}$). 

As can be seen from (\ref{massA}), the decoupling limit is naturally realized 
if $b=\cO(\lambda_i v^2)$ and $t_\beta \gg 1$  or even if  $b \gg \lambda_i v^2$
and  $t_\beta = \cO(1)$. The decoupling regime cannot be realized if 
$b=\lambda_6=\lambda_7=0$. However, this limit is not particularly interesting 
for our purposes since in this case we cannot reach large 
values of $t_\beta$ and, at the same time, be compatible with the LEP 
bounds on $M_{h^0}$.

Finally, let us briefly discuss the possibility of CP-violation, in the simplified 
limit where  $\lambda_3=\lambda_4=\lambda_6=\lambda_7=0$.
In this case only $b$ and $\lambda_5$ can be complex, but we can always 
rotate the Higgs fields such that $b$ is real. The spectrum 
contains a charged Higgs, with mass
\be
M_{H^\pm}^2=\frac{b}{c_\beta s_\beta}-\frac{\text{Re}(\lambda_5)}{2}v^2~,
\ee
and three scalar particles linear combinations of $S_{1,2,3}$. In particular, 
considering the large $t_\beta$ regime, we can expand 
for small $c_\beta$ and obtain at the zeroth order
\bea
M_1^2&\sim&\lambda_2 v^2\,,\\
M_2^2&\sim&\frac{b}{c_\beta}-\frac{v^2}{2}\left(\text{Re}(\lambda_5)-|\lambda_5|\right)\,,\\
M_3^2&\sim&\frac{b}{c_\beta}-\frac{v^2}{2}\left(\text{Re}(\lambda_5)+|\lambda_5|\right)\,,
\eea
from where we can notice the approximate degeneracy of the charged Higgs and the 
two scalars of mass  $M_2$ and $M_3$. Denoting with $h_1,h_2,h_3$ the corresponding 
mass eigenstates, they can be expressed in terms of the original fields $S_1,S_2,S_3$ as
\bea\label{eq:h1}
h_1&\propto& \left(-\frac{2b}{3v^2c_\beta^2},\frac{2}{3}(\lambda_2-\text{Re}(\lambda_5)),\text{Im}(\lambda_5)\right)\,,\\
h_2&\propto&\left(\frac{v^2c_\beta^2}{2b}(\text{Re}(\lambda_5)-3|\lambda_5|+2\lambda_2),1,\frac{\text{Re}(\lambda_5)-|\lambda_5|}{\text{Im}(\lambda_5)}\right)\,,\\\label{eq:h3}
h_3&\propto&\left(\frac{v^2c_\beta^2}{2b}F(\lambda_2,\lambda_5),(\text{Re}(\lambda_5)+|\lambda_5|)^2,\frac{(\text{Re}(\lambda_5)+|\lambda_5|)^3}{\text{Im}(\lambda_5)}\right)\,,
\eea
where the $F(\lambda_2,\lambda_5)$ function has a finite limit for
$\text{Im}(\lambda_5)\to 0$. Two comments are in order: (1)~for
$t_\beta \gg 1$ we obtain the decoupling of $S_1$ from 
$S_{2,3}$, namely we are in the decoupling regime where 
the mass mixing of the two doublets $\Phi_v$ and $\Phi_H$
is negligible; (2)~the mixing of $S_2$ and $S_3$ 
is large even if $t_\beta \gg 1$, and 
vanishes only in the limit where 
$\text{Im}(\lambda_5)\ll\text{Re}(\lambda_5)$, namely in the 
limit of  approximate CP conservation.



\end{document}